\documentclass[longauth]{aa}
\usepackage{graphicx}
\usepackage{natbib}
\usepackage[varg]{txfonts}
\usepackage[usenames,dvips]{color}

\usepackage{enumerate}

\usepackage{subfigure}
\usepackage{float}
\usepackage{xspace}
\usepackage{balance}
\usepackage{upgreek}
\usepackage{array}

\usepackage{hyperref}
\hypersetup{ colorlinks=true,  linkcolor=red, citecolor=blue, urlcolor=blue}

\begin{document}

\title{The VIMOS Public Extragalactic Redshift Survey (VIPERS)
  \thanks{Based on observations collected at the European Southern
    Observatory, Cerro Paranal, Chile, using the Very Large Telescope
    under programmes 182.A-0886 and partly 070.A-9007.  Also based on
    observations obtained with MegaPrime/MegaCam, a joint project of
    CFHT and CEA/DAPNIA, at the Canada-France-Hawaii Telescope (CFHT),
    which is operated by the National Research Council (NRC) of
    Canada, the Institut National des Sciences de l'Univers of the
    Centre National de la Recherche Scientifique (CNRS) of France, and
    the University of Hawaii. This work is based in part on data
    products produced at TERAPIX and the Canadian Astronomy Data
    Centre as part of the Canada-France-Hawaii Telescope Legacy
    Survey, a collaborative project of NRC and CNRS.}}
\subtitle{Environmental effects shaping the galaxy stellar mass function}

\author{
I.~Davidzon\inst{1,2}
\and O.~Cucciati\inst{3,2}
\and M.~Bolzonella\inst{2}
\and G.~De Lucia\inst{4}
\and G.~Zamorani\inst{2}
\and S.~Arnouts\inst{1,5}
\and T.~Moutard\inst{1} 
\and O.~Ilbert\inst{1}
\and B.~Garilli\inst{6}
\and M.~Scodeggio\inst{6}
\and L.~Guzzo\inst{7,8}
\and U.~Abbas\inst{9}
\and C.~Adami\inst{1}
\and J.~Bel\inst{7,10,11}       
\and D.~Bottini\inst{6}
\and E.~Branchini\inst{12,13,14}
\and A.~Cappi\inst{2,15}
\and J.~Coupon\inst{16}          
\and S.~de la Torre\inst{1}
\and C.~Di Porto\inst{2} 
\and A.~Fritz\inst{6}
\and P.~Franzetti\inst{6}
\and M.~Fumana\inst{6}     
\and B.~R.~Granett\inst{7}
\and L.~Guennou\inst{17,1}
\and A.~Iovino\inst{7}
\and J.~Krywult\inst{18}
\and V.~Le Brun\inst{1}
\and O.~Le F\`evre\inst{1}
\and D.~Maccagni\inst{6}
\and K.~Ma{\l}ek\inst{19}
\and F.~Marulli\inst{3,2,20}
\and H.~J.~McCracken\inst{21}
\and Y.~Mellier\inst{21}
\and L.~Moscardini\inst{3,2,20}
\and M.~Polletta\inst{6}
\and A.~Pollo\inst{22,19} 
\and L.~A.~M.~Tasca\inst{1}
\and R.~Tojeiro\inst{23}
\and D.~Vergani\inst{24,2}
\and A.~Zanichelli\inst{25}
}

\offprints{\texttt{iary.davidzon@lam.fr}}  
  
\institute{
Aix Marseille Universit\'e, CNRS, LAM (Laboratoire d'Astrophysique de Marseille) UMR 7326, 13388, Marseille, France  
\and INAF - Osservatorio Astronomico di Bologna, via Ranzani 1, I-40127, Bologna, Italy 
\and Dipartimento di Fisica e Astronomia - Alma Mater Studiorum Universit\`{a} di Bologna, viale Berti Pichat 6/2, 40127 Bologna, Italy 
\and INAF - Osservatorio Astronomico di Trieste, via G. B. Tiepolo 11, 34143 Trieste, Italy 
\and Canada-France-Hawaii Telescope, 65--1238 Mamalahoa Highway, Kamuela, HI 96743, USA 
\and INAF - Istituto di Astrofisica Spaziale e Fisica Cosmica Milano, via Bassini 15, 20133 Milano, Italy 
\and INAF - Osservatorio Astronomico di Brera, Via Brera 28, 20122 Milano, via E. Bianchi 46, 23807 Merate, Italy 
\and Dipartimento di Fisica, Universit\`a di Milano-Bicocca, P.zza della Scienza 3, I-20126 Milano, Italy 
\and INAF - Osservatorio Astronomico di Torino, 10025 Pino Torinese, Italy  
\and Universit\'e de Toulon, CNRS, CPT, UMR 7332, 83957 La Garde, France 
\and Aix Marseille Universit\'e, CNRS, CPT, UMR 7332, 13288 Marseille, France 
\and Dipartimento di Matematica e Fisica, Universit\`{a} degli Studi Roma Tre, via della Vasca Navale 84, 00146 Roma, Italy 
\and INFN, Sezione di Roma Tre, via della Vasca Navale 84, I-00146 Roma, Italy 
\and INAF - Osservatorio Astronomico di Roma, via Frascati 33, I-00040 Monte Porzio Catone (RM), Italy 
\and Laboratoire Lagrange, UMR7293, Universit\'{e} de Nice Sophia Antipolis, CNRS, Observatoire de la C\^ote d’Azur, 06300 Nice, France 
\and Astronomical Observatory of the University of Geneva, ch. d'Ecogia  16, 1290 Versoix, Switzerland
\and Institut d'Astrophysique Spatiale, Universit\'{e} Paris-Sud, CNRS, UMR8617, 91405, Orsay, France 
\and Institute of Physics, Jan Kochanowski University, ul. Swietokrzyska 15, 25-406 Kielce, Poland 
\and National Centre for Nuclear Research, ul. Hoza 69, 00-681 Warszawa, Poland 
\and  INFN, Sezione di Bologna, viale Berti Pichat 6/2, I-40127 Bologna, Italy 
\and Institute d'Astrophysique de Paris, UMR7095 CNRS, Universit\'{e} Pierre et Marie Curie, 98 bis Boulevard Arago, 75014 Paris, France 
\and Astronomical Observatory of the Jagiellonian University, Orla 171, 30-001 Cracow, Poland 
\and Institute of Cosmology and Gravitation, Dennis Sciama Building, University of Portsmouth, Burnaby Road, Portsmouth, PO1 3FX 
\and INAF - Istituto di Astrofisica Spaziale e Fisica Cosmica Bologna, via Gobetti 101, I-40129 Bologna, Italy 
\and INAF - Istituto di Radioastronomia, via Gobetti 101, I-40129, Bologna, Italy 
}

\titlerunning{VIPERS -- Environmental effects shaping the GSMF.}
\authorrunning{I.~Davidzon et al.}

\date{Received ... / Accepted ...}

\abstract{We exploit the first public data release of VIPERS to investigate 
  environmental effects in the evolution of galaxies 
  between $z\sim0.5$ and $0.9$. 
  The large number of spectroscopic redshifts (more than 50\ 000) 
  over an area of about $10\,\mathrm{deg}^2$ provides a galaxy sample 
  with high statistical power.   
  The  accurate redshift measurements ($\sigma_z = 0.00047(1+z_\mathrm{spec})$) 
  allow us to robustly isolate galaxies living in the lowest-  and highest-density environments 
  ($\delta<0.7$ and $\delta>4$, respectively) 
  as defined in terms of spatial 3D density contrast $\delta$. 
  We estimate the stellar mass function of galaxies residing in these two environments, and 
  constrain the high-mass end ($\mathcal{M}\gtrsim 10^{11}\,\mathcal{M}_\odot$)  
  with unprecedented precision. 
  We find that the galaxy stellar mass function  in the densest regions has a different shape 
  than that measured at low densities, with an enhancement of massive galaxies  
  and a hint of a flatter (less negative) 
  slope  at $z<0.8$. 
  We   normalise each mass function to the comoving volume 
  occupied by the corresponding  environment, and  relate  
  estimates from different redshift bins. We observe an evolution of the stellar mass function 
  of VIPERS galaxies in high densities, while the low-density one is nearly constant.  
  We compare these results to semi-analytical models and find consistent environmental 
  signatures in the simulated stellar mass functions. 
  We discuss how the halo mass function and fraction of central/satellite galaxies depend 
  on the environments considered, 
  making intrinsic and environmental properties of galaxies physically coupled, 
  and therefore difficult to disentangle.  
  The evolution of our low-density regions is well described by the formalism introduced 
  by Peng et al.~(2010), and is consistent with the idea that galaxies become progressively 
  passive because of internal physical processes. 
  The same formalism could also describe the evolution of the mass function in the 
  high density regions, but only if a significant contribution from dry mergers is considered. 
}


\maketitle

\newpage 

\section{Introduction}
\label{Introduction}

  After several decades from the pioneering work  
  of~\citet[][]{Oemler1974,Davis&Geller1976,Sandage&Visvanathan1976}, 
  the role of environment in driving galaxy evolution still represents a  research 
  frontier.
  Several correlations have been observed between the 
  place in which galaxies reside and their own properties 
  \citep[see e.g.][for a review]{Blanton2009}, 
  but the mechanisms responsible for them remain 
  poorly understood. Even the well-established morphology-density relation 
  \citep{Dressler1980,Postman&Geller1984} has a number of 
  contrasting interpretations 
  \citep[cf][]{Thomas2010,vanderWel2010,Cappellari2011}.

  Many pieces of evidence  
  suggest that the environment
  has a fundamental influence \citep[e.g.][]{ 
  Cooper2008,Cucciati2010,Burton2013}. 
  In particular, some 
  of the processes halting the production of new stars 
  (the so-called ``galaxy quenching'') should be related to the 
  dense intergalactic medium (e.g., ram pressure stripping) 
  and/or interactions with nearby galaxies \citep[for more details, 
  see e.g.][]{Boselli&Gavazzi2006,Gabor2010,Woo2012}.
  On the contrary, 
  other authors  consider the galaxy stellar mass  ($\mathcal{M}$) 
  or the halo mass ($\mathcal{M}_\mathrm{h}$) 
  the main evolutionary drivers, 
  with a secondary -- or even negligible -- 
  contribution by their environment \citep[e.g.][]{ 
  Pasquali2009,Thomas2010,Grutzbauch2011}.
    
  Classical discussions contrast a scenario in which the 
  fate of a galaxy is determined primarily by physical processes 
  coming into play after the galaxy has become part of a group or 
  of a cluster (``nurture''), to one in which the observed environmental 
  trends are established before these events, 
  and primarily determined by internal physical processes (``nature''). 
  However, this dichotomy is simplistic,
  as stellar mass and environment are inter-related. 
  In fact, the parametrisation of the latter is 
  often connected to the gravitational  
  mass of the hosting halo, which is also physically 
  coupled to galaxy stellar mass. 
   Therefore,  in a scenario of 
   hierarchical accretion,  it is expected that  
  most massive galaxies show a correlation with overdensities  
  \citep{Kauffmann2004,Abbas2005,Scodeggio2009}.
  For this reason,
  it is misleading
  to contrast stellar mass and 
  environment as two separate aspects of galaxy evolution 
  \citep[see the discussion in][]{DeLucia2012}.  
  
  Another crucial point is how the environment is defined. 
  One possibility is to identify high-density regions as
  galaxy groups and clusters, in contrast to a
  low-density  ``field'', sometimes ambiguously defined.
  When halo mass estimates are used, the  
  classification is more tightly related to the underlying distribution
  of dark matter, with galaxies often divided in satellite and central
  objects  \citep[][]{vandenBosch2008a}. 
  Other methods, involving galaxy counts, 
  can identify a broad range of densities 
  with a resolution 
  from a few Megaparsecs down to $\sim100\,\mathrm{kpc}$;
   they are based on the two-point clustering \citep[e.g.][]{Abbas2005}, 
  Voronoi's tessellation \citep[e.g.][]{Marinoni2002},  
  or  the computation of the galaxy number density inside a 
  window function, which is the approach used in the present work.  
  In general, different methods probe galaxy surroundings on 
  different scales 
  \citep[][]{Muldrew2012}. Nonetheless, the method we adopt here 
  (based on the fifth nearest neighbourhood) 
  is expected to be in overall good agreement with other 
  robust estimators as Delaunay's or Voronoi's tessellations 
  \citep[see][]{Darvish2015}.

  In this kind of research, 
   the galaxy stellar mass function 
  (GSMF) is one of the most effective tools. 
  Especially when computed inside a specific 
  environment, the GSMF requires excellent data,  
  derived from the observation of wide fields 
  or a large number of clusters/groups. 
  For this reason, only few studies on the GSMF 
  consider the environmental dependence aspects 
  \citep[e.g.][]{Baldry2006,Bundy2006,Bolzonella2010,Vulcani2012,
  Giodini2012,Annunziatella2014,Hahn2015,Mortlock2015}.
  Although still fragmentary, 
  an interesting picture is emerging from these pieces of work. 
    In the local Universe, \citet[][SDSS data]{Baldry2006} 
  observe a  correlation between the turn-off mass of the 
  GSMF ($\mathcal{M_\star}$) 
  and the local density (which they estimate
  as an average between the fourth and fifth nearest neighbour). 
  In the lowest densities, they estimate   
  $\log(\mathcal{M_\star/M_\odot})\simeq10.6$, reaching values of   
  about 11.0 in the densest environment.  
  Probing a larger redshift range (from $z\sim1$ to $\sim0.1$)  
  \citet{Bolzonella2010}  detect 
   environmental effects for zCOSMOS \citep[][]{Lilly2007} galaxies: 
  the passive population grows  more rapidly
  inside regions of high density (recovered by counting the fifth nearest neighbour 
  of each galaxy). 
  The authors find 
  this trend by studying the redshift evolution 
  of $\mathcal{M}_\mathrm{cross}$, 
  i.e.~the value of stellar mass at which the active and passive GSMFs 
  intersect each other 
  \citep[see also][]{Bundy2006,Peng2010,Annunziatella2014}. 
  Recent studies indicate that  already at $z\sim1$ 
   the assembly of massive galaxies is faster 
  in overdensities \citep{Mortlock2015}.  
    Using a slightly different classification with respect to 
    \citeauthor{Bolzonella2010}, i.e.~relying on the third nearest
  neighbour, \citet{Bundy2006} seek for environmental
  effects in the stellar mass function of DEEP2 galaxies, 
  from $z=0.4$ to 1.4. 
  The evolution they find shows a mild 
  dependence on local environment,
  such that \citeauthor{Bundy2006} quantify it 
  as a secondary driver with respect to stellar mass.  
  Other studies compare the  stellar mass functions
  of clusters, groups, and isolated (or ``field'') galaxies.
  \citet{Kovac2010b}, using the 10k zCOSMOS sample,
  confirm the trend noted by \citet{Bolzonella2010}:
  massive galaxies preferentially reside inside groups. 
  \citet{Annunziatella2014} find that the passive galaxy stellar 
  mass function  
   in a cluster of the CLASH-VLT survey 
   has a steeper slope 
   in the core of the cluster than in the 
  outskirts \citep[see also][]{Annunziatella2015}. 
  On the other hand, 
  \citet{Calvi2013} and \citet{Vulcani2012,Vulcani2013}
  compare galaxy clusters and general field up to 
  $z\simeq0.8$, 
  without detecting any significant difference 
  in the respective GSMFs. Also 
 \citet{vanderBurg2013}  
   find similar shapes for active/passive mass functions  
  in each environment, although 
  the total GSMFs differ from each other because of 
  the different percentage of passive galaxies in  their   
  clusters  at $0.86<z<1.34$ 
  with respect to  the field.  
  We note however that these analyses are based on 
  different kinds of datasets: \citeauthor{Vulcani2013} and 
  \citeauthor{vanderBurg2013} use samples of several 
  clusters, while \citeauthor{Annunziatella2014} focus on 
  one system but with deeper and wider observations. 

  We aim to provide new clues in this context, exploiting the 
  VIMOS Public Extragalactic Redshift Survey 
  \citep[VIPERS,][]{Guzzo2014} to search for environmental 
  effects between $z\simeq1$ and $z\simeq0.5$. 
  As shown in a previous paper of this series 
  \citep[][hereafter D13]{Davidzon2013}, the VIPERS data allow 
  robust measurement of the GSMF. 
  The accurate VIPERS redshifts are also the cornerstone to estimate  
  the local density contrast around each galaxy;
  this task has been carried out 
  in \citet[][]{Cucciati2014a} and will be further 
  developed in another study focused on the colour-density relation 
  (Cucciati et al., in prep.). 
  In the present paper, Sect.~\ref{Data}
  contains a general description of the survey.  
  The computation of local density contrast  
  is summarised in the same Section, along with 
  the derivation of other fundamental galaxy quantities 
  (in particular galaxy stellar mass).
  In Sect.~\ref{Classifications} we present our classification of
  environment and  galaxy types.
  After posing these definitions, we are able to estimate the GSMF
  in low- and high-density regions of VIPERS, also considering
  the passive and active galaxy samples separately 
  (Sect.~\ref{Gsmf results}).
  The interpretation of our results is discussed in Sect.~\ref{Discussion},
  while conclusions are in Sect.~\ref{Conclusions}.
  
  Our measurements assumes a flat $\Lambda$CDM cosmology in which 
  $\Omega_m = 0.25$, $\Omega_\Lambda= 0.75$, and $h_{70}=H_0 /
  (70\,\mathrm{km\,s^{-1}\,Mpc^{-1}})$, unless specified otherwise.
  Magnitudes are in the AB system \citep{Oke1974}.

  \section{Data}
  \label{Data}
    
  Since 2008, VIPERS has probed 
  a volume of $\sim1.5 \times 10^8\,\mathrm{Mpc}^3\,h_{70}^{-3}$
  between $z=0.5$ and $1.2$,  providing  the largest spectroscopic galaxy
  catalogue at intermediate redshifts.
  The final public release, expected in 2016, will cover $24$\,deg$^2$,
  including about 90\,000 galaxies and
  AGNs in the  magnitude range of $17.5\leqslant i \leqslant 22.5$.
  The first public data release (PDR1),
  consisting of 57\,204 spectroscopic measurements,  
  has been presented in \citet{Garilli2014} and is now
  available on the survey database\footnote{\url{http://vipers.inaf.it/rel-pdr1.html}}.

  From a cosmological perspective, the main goals of VIPERS is  
  measuring the growth rate of structure \citep{delaTorre2013b}. 
   Additional science drivers refer also to extragalactic research fields, 
  to investigate  a wide range of galaxy properties at an epoch when
  the Universe was about half its current age
  (\citealp{Marchetti2013,Malek2013}; D13; \citealp{Fritz2014}). 
  In addition, in the context of the present study, it is worth mentioning the VIPERS 
  papers   that describe  the relation between baryons and dark matter
  through the galaxy bias factor \citep{Marulli2013,DiPorto2014,Cappi2015,Granett2015}. 
  Both the galaxy density field and the galaxy bias, if the latter is measured  
  as a function of stellar mass and/or luminosity, are intimately linked to  clustering  
  and the total matter distribution.
  We refer the reader to \citet{Guzzo2014} and \citet{Garilli2014}
  for further details on the survey construction and the
  scientific investigations being carried out by the VIPERS collaboration.

\subsection{Photometry}
\label{Photometry}

  The spectroscopic survey is 
  associated with photometric ancillary data
  obtained from both public surveys and dedicated observations.
  The VIPERS targets have been selected within two fields of the
  Canada-France-Hawaii Telescope Legacy Survey Wide
  (CFHTLS-Wide\footnote{\url{http://www.cfht.hawaii.edu/Science/CFHLS/}}),
  namely W1 and W4. 
  The CFHTLS optical magnitudes were derived by the Terapix
  team\footnote{Data available at \url{http://www.terapix.iap.fr}  
  (T0005 data release).} 
  by means of \texttt{SExtractor} 
  \citep[MAG\_AUTO in double image mode, see][]{Bertin&Arnouts1996} 
  in the filters  
  $u^\ast$, $g^\prime$, $r^\prime$, $i^\prime$, and $z^\prime$. 
  Photometric redshifts ($z_\mathrm{phot}$) have been estimated by using
  such magnitudes, following the procedure described in \citet{Coupon2009};
  their uncertainty is $\sigma_\mathrm{zphot}=0.035(1+z_\mathrm{phot})$.
  This photometric catalogue  is limited at $i\leqslant22.5$, and we refer to it as 
  the ``parent sample'' of VIPERS.
  Sources whose  
  quality was deemed insufficient 
  for our analysis (e.g.~because of nearby stars) 
  have been excluded by means of angular masks.

  Beyond optical data, a $K_s$-band follow-up 
  added information in the near-infrared (NIR) range.  
  Data were collected 
  by means of the WIRCam instrument at CFHT, 
  setting an optimised depth 
  to match the brightness of the spectroscopic sources 
  (Moutard et al., in prep.). 
  These observations cover almost completely the W4 field, 
  while in W1 a $1.6\times0.9$\,deg$^2$ area is missing
  (see D13, Fig.~1).
  At $K\leqslant22.0$ (that is the $5\sigma$ limiting magnitude), 
  96\% of the VIPERS objects in W4 have a 
  WIRCam counterpart, as compared to
  80\% in W1. 
  When estimating galaxy stellar masses by fitting galaxy
  spectral energy distributions (SEDs),
  NIR photometry can be critical, 
  e.g.~to avoid age underestimates
  \citep[see][]{Lee2009}.
  For this reason, $K_\mathrm{WIRCam}$ has been
  complemented by the UKIDSS data\footnote{\url{http://www.ukidss.org},
  note that Petrosian magnitudes in the database are in Vega system,
  but conversion factors to AB system are provided 
  by the UKIDSS team on the reference website.
  }.
  The sky region that WIRCam did not observe in W1
  is fully covered by the UKIDSS-DXS survey, 
  which has been used also in W4 
  -- together with the shallower UKIDSS-LAS -- 
  for less than $300$ sparse objects not matched 
  with $K_\mathrm{WIRCam}$. 
  After that, the fraction of our spectroscopic sample having
  $K$-band magnitude rises to $97$\% both in W1 and in W4;
  in absence of $K$ magnitudes, we use (where possible) 
  the $J$ band  from UKIDSS.
  The comparison between $K_\mathrm{WIRCam}$ and 
  $K_\mathrm{UKIDSS}$  was
  performed in D13: the two surveys are in
  good agreement, and we can safely combine them.
  
  In addition, about 32\% of the spectroscopic targets in W1
  lie in the XMM-LSS field and have been associated with
  infrared (IR) sources observed by SWIRE\footnote{\url{http://swire.ipac.caltech.edu/swire}}.
  Since our SED fitting  (Sect.~\ref{Sed fitting}) 
  uses models of stellar population synthesis 
  that do not reproduce the re-emission from dust, 
  we only considered magnitudes in the $3.6\,\mu$m and
  $4.5\,\mu$m bands (it should be also noticed that 
  beyond those wavelengths SWIRE  is
  shallower, and source detection is very sparse).

  \subsection{Spectroscopy}
  \label{Dataspec}
  
  We extract our galaxy sample from  the 
  same spectroscopic catalogue used  
  in D13.
  That catalogue includes 53\,608 galaxy spectra
  with $i \leqslant i_\mathrm{lim} \equiv22.5$. 
  Along with the limiting magnitude, 
  an additional criterion for target selection,
  based on $(g-r)$ and $(r-i)$ colours,
  was successfully applied 
  to enhance the probability of observing 
  galaxies at $z>0.5$ \citep[see][]{Guzzo2014}.
  
  Spectra were observed at VLT using the VIMOS multi-object
  spectrograph \citep{LeFevre2003a} 
  with the LR-Red  grism ($R=210$) 
  in a wavelength range of 5500--9500\,\AA.
  The four quadrants of the VIMOS instrument, separated 
  by gaps $2\arcmin$ wide, produce a characteristic 
  footprint that we have accounted for 
  by means of spectroscopic masks. 
  Besides gaps, a few quadrants are missing
  in the survey layout \citep[][Fig.~10]{Guzzo2014}
  because of  technical issues in the spectrograph set-up. 
  After removing the vignetted parts of each pointing,
  the effective area covered by the survey is 
  5.34 and 4.97\,deg$^2$, in W1 and W4 respectively.
  To maximise the number of targets, 
  we used short slits as proposed in \citet{Scodeggio2009}.
  As a results we targeted $\sim45\%$  of available sources  
  in a single pass.
  
  A description of the VIPERS data reduction  
  can be found in \citet{Garilli2012a}. At the end of
  the pipeline, a validation process was carried out
  by team members, who checked 
  the measured redshifts  
  and assigned a quality flag ($z_\mathrm{flag}$) to each of them.
  Such a flag corresponds to the confidence level (CL) of
  the measurement, according to the same scheme 
  adopted by previous surveys like
  VVDS \citep{LeFevre2005} and zCOSMOS.
  The sample we will use 
  includes  galaxies with
  $2 \leqslant z_\mathrm{flag} \leqslant 9$, corresponding to 
  95\% CL. 
  It comprises 34\,571  spectroscopic measurements  
  between $z=0.5$ and  0.9, i.e.~the redshift range of the present analysis. 
  We estimate the error in the $z_\mathrm{spec}$ measurements 
  from repeated observations. It is  
   $\sigma_z = 0.00047(1+z_\mathrm{spec})$, 
  corresponding to a velocity uncertainty of $\sim140\,\mathrm{km\,s}^{-1}$ 
  \citep[][]{Guzzo2014}.
  We provide each object with a statistical weight $w(i,z)$
  to make this sample representative of all the photometric galaxies at
  $i<22.5$ in the survey volume.
  We estimate weights by considering three 
  selection functions: the target sampling rate (TSR),
  the spectroscopic success rate (SSR), and the
  colour sampling rate (CSR). 
  Further details about TSR, SSR, and CSR 
  are provided in \citet{Fritz2014}, \citet{Guzzo2014}, 
  and \citet{Garilli2014}. The overall sampling rate, 
  i.e.~$\mathrm{TSR}\times\mathrm{SSR}\times\mathrm{CSR}$ 
  is on average 35\%.

  \subsection{SED fitting estimates}  
  \label{Sed fitting}

  We estimate several quantities, 
  in particular galaxy stellar masses and absolute magnitudes,
  by means of SED fitting, 
  in a similar way to D13.
  Through this technique, physical 
  properties of a given galaxy can be  derived
  from the template 
  (i.e., the synthetic SED) that
  best reproduces its multi-band
  photometry (after redshifting the template 
   to $z=z_\mathrm{spec}$ or $z_\mathrm{phot}$). 
  To this purpose, we use the 
  code \textit{Hyperzmass}, a modified version of 
  \textit{Hyperz} \citep{Bolzonella2000} developed by
  \citet{Bolzonella2010}.
  The software selects the best-fit template as the one that
  minimises the $\chi^2$.
 
  To build our library of galaxy templates we 
  start from the simple stellar populations  
  modelled by  \citet[][hereafter BC03]{Bruzual2003}.
  The BC03 model assumes a universal initial mass function (IMF)
  and a single non-evolving metallicity ($Z$)
  for the stars belonging to a given simple stellar population (SSP).
  Many SSPs are combined and integrated in order to reproduce
  a galaxy SED.

  As in D13, 
  we choose SSPs with \citet{Chabrier2003} IMF, 
  having metallicity either
  $Z=Z_\odot$ or $Z=0.2Z_\odot$ 
  to sample the metallicity range observed  
  for galaxies at $z\sim0.8$ \citep{Zahid2011}.
  We adopt only two values to limit 
  degeneracy with other parameters such as the age. 
  Synthetic galaxy SEDs are derived 
  by evolving the SSPs in agreement with a given
  star formation history (SFH). 
  We assume eleven SFHs: one with a constant SFR,
  and ten having an exponentially declining 
  profile, i.e.~$\mathrm{SFR}\propto\exp(-t/\tau)$ 
  with values of $\tau$ ranging 
  between 0.1 and $30\,\mathrm{Gyr}$.
  The formation redshift of our galaxy templates
  is not fixed, but the ages allowed in the fitting procedure
  range from 0.09\,Gyr to the age of the Universe at
  the redshift of the fitted galaxy. 
  Composite SFHs could be considered
  by adding random bursts of star formation to
  the exponential (or constant) SFR, as in 
  \citet{Kauffmann2003c}. 
  However, in D13 we
  checked that replacing smooth SFHs with
  more complex ones has a critical impact on the 
  stellar mass estimate (i.e., more than 
  0.3\,dex difference) 
  only for a small fraction ($<10$\%) of the VIPERS
  galaxies, while for the majority of the sample the
  change is less than $\sim0.1$\,dex
  \citep[see also][]{Pozzetti2007}.
  Similar conclusions are drawn by \citet{Mitchell2013},
  who find that
  the exponential decrease  
  is a reasonable approximation
  of the true (i.e., composite) SFH of their simulated galaxies: their
  SED fitting estimates show small scatter and no systematics
  with respect to the stellar masses obtained from the
  theoretical model
  \citep[see also][whose results do not change significantly 
  when using either composite or ``delayed'' SFHs]{Ilbert2013}.
  
  Attenuation by dust is modelled
  by assuming either  \citet{Calzetti2000} or Pr\'evot-Bouchet
  \citep{Prevot1984,Bouchet1985} extinction law. 
  For both, we let the $V$-band attenuation 
  vary from $A_V=0$ (i.e., no dust) to 3\,mag, with steps of 0.1.
  No prior is implemented to discriminate between the two 
  extinction laws: 
  for each galaxy the model is chosen that 
  minimises the $\chi^2$.
  We exclude from our library those
  templates having an unphysical SED, 
  according to observational evidence. 
  Galaxy models with $\mathrm{age}/\tau > 4$ and $A_V > 0.6$ are not 
  used in the fitting procedure, since they represent old galaxies 
  with an excessive amount of dust  compared to what observed 
  in the local universe 
  \citep[cf][Fig.~3]{Brammer2009}.
  We also rule out best-fit solutions 
  representing early-type 
  galaxies (ETGs) with too young ages, i.e.~models   
  with $\tau \leqslant 0.6\,\mathrm{Gyr}$   
   and redshift of formation $z_\mathrm{form}< 1$ 
  \citep[evidence of high $z_\mathrm{form}$ of ETGs can be found e.g.~in][]{Fontana2004,
  Thomas2010}. Any other combination of parameters within the ranges 
  mentioned above is allowed.
  Considering all these parameters and their allowed ranges,  
  our SED fitting should provide us with 
  stellar mass estimates with
  an uncertainty of  $\lesssim0.3$\,dex, according to 
  \citet{Conroy2009b}.  
  Moreover, we emphasise that the lack of IR photometry 
  for a small part of the VIPERS sample (see Sect.~\ref{Photometry}) 
  does not introduce significant bias, as already tested in
  D13.
  
  In addition to stellar mass,
  we estimate  absolute magnitudes 
  in several  bands
  from the same best-fit SED. 
  To minimise the model dependency, 
  we take the apparent magnitude 
  in the closest filter to
  the  rest-frame wavelength of interest, 
  and apply a k- and colour-correction 
  based on the SED shape.
  In this way, the outcome is little sensitive
  to the chosen template, relying mainly
  on the observations. Typical uncertainties of this
  procedure, when applied to optical/NIR filters, 
  are about 0.05\,mag at  $0.5<z<0.9$ 
  \citep[][]{Fritz2014}.

 \subsection{Galaxy density contrast}  
 \label{Density contrast}

  To characterise the different environments in which galaxies
  live (Sect.~\ref{Environment definition}) we rely on
  the galaxy density contrast ($\delta$). This quantity is related to the local
  concentration of galaxies (i.e.~the galaxy density field $\rho$) and the mean
  galaxy density ($\bar{\rho}$) such that
  $\delta=(\rho-\bar{\rho})/\bar{\rho}$.
  Although $\rho$ is a point field
  indirectly connected to matter density, 
  it is a good proxy of the 
  underlying matter distribution: through various smoothing schemes 
  (included the one described here) it is possible to recover the latter
  from the former with a scale-independent bias factor
  \citep{Amara2012,DiPorto2014}. 
  The procedure adopted here is thoroughly described in a 
  companion paper \citep{Cucciati2014a}.

  To derive the local density of a given galaxy,
  we count objects inside a filter centred on it. 
  Those objects that trace $\rho$ are part of
  a ``volume-limited'' sample that includes
  both spectroscopic
  and photometric galaxies. The latter ones come from
  the photometric parent catalogue, which contains CFHTLS 
  sources with the same $i$-band cut of VIPERS 
  (see Sect.~\ref{Photometry}).
  To build such a sample, we select galaxies in W1 and W4
  with  $M_B < -20.4 -Qz$. The factor
  $Q$ takes into account the evolution in redshift of the 
  characteristic galaxy luminosity, as determined by   
  $M_B^\star$ in the galaxy luminosity 
  functions 
  \citep[see more details e.g.~in][]{Moustakas2013}. 
  We set 
  $Q=1$ according to the zCOSMOS luminosity function
  \citep[which encompasses $z\sim0.2$ to 0.9, see][]{Zucca2009}. 
  The volume-limited sample is complete up to $z=0.9$, 
  and traces the underlying cosmic structure
  avoiding strongly evolving 
  bias that instead a flux-limited sample would  produce 
  \citep[cf][]{Amara2012}.
  We will refer to this volume-limited sample 
   as the sample of
  ``tracers'' (to be distinguished from 
   the VIPERS sample for 
   which we will compute $\delta$).

  Among those tracers, 14\,028 objects have a $z_\mathrm{spec}$ 
  with $z_\mathrm{flag}\in[2,9]$ while more than $100\,000$ have 
  a $z_\mathrm{phot}$.  
  The large number of spectroscopic redshifts -- and
  their accuracy -- is crucial to robustly determine the 
  density field in the 3-dimensional (redshift) volume:  
  generally, when using 
  photometric redshifts only, the reconstruction along the 
  line of sight is prevented by their larger photo-$z$ errors 
  \citep[e.g.][]{Cooper2005,Scoville2013}.  
  We will compute $\delta$ for galaxies beyond $z=0.51$,
  to avoid the steep decrease of $N(z)$ 
  at $z \lesssim 0.5$ \citep[see][Fig.~13]{Guzzo2014}
  that  could affect our density estimates.

  Thanks to the photometric redshifts, there is a sufficient number of  
  (photometric) tracers
  also in the gaps produced by the footprint of VIMOS and
  in the missing quadrants.\footnote{
  Nevertheless, \citet{Cucciati2014a} demonstrate that
  the major source of uncertainties in the procedure
  is not the presence of  gaps 
  but the incompleteness of the spectroscopic sample
  (i.e.~the $\sim35$\% sampling rate). Besides that, we emphasise that 
  the  $z_\mathrm{phot}$ sample is crucial to avoid 
  any environmental bias caused by the slit assignment. 
  In fact, the VIMOS sampling rate could be slightly smaller 
  in crowded regions, because of the minimum distance required 
  between two nearby slits. } 
  In absence of a secure spectroscopic measurement,
  we apply a modified version of the method described
  by \citet{Kovac2010a}. The key idea of the method 
  is that  galaxy clustering along 
  the line of sight, 
  recovered by using spectroscopic redshifts,   
  provides information about the  radial positions  
  of a photometric object, i.e.~it is likely to lie 
  where the clustering is higher. 
  Thus,  to each photometric tracer
  we assign a distribution of $z_\mathrm{peak}$ values,
  together with an ensemble of
  statistical weights.
  For each value of $z_\mathrm{peak}$, 
  the associated weight $w_\mathrm{peak}$ represents 
  the relative probability for the object to be 
  at that given redshift
  (the sum of weights is normalised to unity).
  In other words,  
  the $z_\mathrm{peak}$ values are the ``most likely''
  radial positions of a photometric tracer.
  
  In detail, to determine $z_\mathrm{peak}$ and $w_\mathrm{peak}$, 
  we proceed as follows.
  We start from the probability distribution function (PDF) of 
  the measured
  $z_\mathrm{phot}$, assumed to be
  a Gaussian 
  with rms equal to $\sigma_\mathrm{zphot}$.
  We also determine  $N(z)$, that is 
  the galaxy distribution along the line of sight 
  of the target. To do that, we take  
  all the objects of the spectroscopic sample lying 
  inside a cylinder with  
  $7.1\,h_{70}^{-1}\,\mathrm{Mpc}$ comoving radius\footnote{ 
  This value  corresponds to a radius equal to
  $5\,\mathrm{Mpc}$ if one assumes 
  $H_0=100\,\mathrm{km\,s^{-1}\,Mpc^{-1}}$ \citep[as in][]{Cucciati2014a}
  } 
  and half-depth of $\pm3\sigma_\mathrm{zphot}$; the cylinder is
  centred in the coordinates (RA, Dec, $z_\mathrm{phot}$) of the
  considered galaxy. The 
  desired $N(z)$ distribution is obtained from those objects, 
  using their $z_\mathrm{spec}$ values (without errors) 
  in bins of $\Delta z =0.003$.
  Then, we multiply the PDF of $z_\mathrm{phot}$ 
  by $N(z)$ and renormalise the resulting function. 
  In this way we obtain a new PDF whose peaks 
  represent the desired set of
  $z_\mathrm{peak}$ values. Their respective $w_\mathrm{peak}$
  are provided according to the relative height of each peak 
  (the sum of them being equal to one).

  Thus, we compute the local density $\rho$ 
  for each of the 33\,952 galaxies of the VIPERS 
  (flux-limited) catalogue from $z=0.51$ to 0.9.
  Given the galaxy coordinates  
  $r_g=(\mathrm{RA}_g,\mathrm{Dec}_g)$
  and redshift $z_g$,
  $\rho(r_g,R_{\mathrm{5NN}})$  
  is equal to the number of  tracers 
  inside a cylindrical filter centred in $r_g$;
  the cylinder has half-depth 
  $\Delta v = \pm 1000\,\mathrm{km}\,\mathrm{s}^{-1}$ 
  and radius equal to $R_{\mathrm{5NN}}$, i.e.~the projected distance of 
  the fifth closest tracer 
  (or fifth nearest neighbour, hereafter 5NN). 
  It should be noticed that 
  such an estimate depends on the absolute magnitude of 
  the tracers. By using fainter tracers  
  (e.g., limited at $M_B<-19.5-z$) the object identified as 
  5NN would change and $R_\mathrm{5NN}$ would be generally smaller. 
  However, although the absolute value of $\delta$ 
  varies as a function of tracer luminosity,  
  we are interested in a relative classification 
  that divides galaxies in  
  under- and over-densities (see Sect.~\ref{Environment definition}). 
  Therefore, a different cut in $M_B$  
  would not alter our findings, as we verified that 
  the galaxy ranking in density contrast 
  would be on average preserved.   
  On the other hand, fainter tracers 
  would be incomplete already at  
  lower redshifts (for example by assuming $M_B<-19.5-z$ 
  we would restrict our analysis at $z<0.7$).

  The density contrast is defined 
  on scales that   
  differs 
  from one galaxy to another. Namely, in our reconstruction $R_\mathrm{5NN}$ 
  ranges from $\sim\!2.8$ to $8.6\,h_{70}^{-1}\,\mathrm{Mpc}$ 
  while moving from the densest regions toward galaxies 
  with the lowest $\rho$. 
  Probing a non-uniform scale
  does not impair our analysis, because we are interested in a
  relative classification of different environments
  (see Sect.~\ref{Environment definition}).  
  The 5NN estimator leads to the desired ranking.
  We adopt the 5NN because 
  it is an adaptive estimator that efficiently samples 
  a broad range  of densities. 
  Using, instead, a fixed radius of
  $\sim\!3\,h_{70}^{-1}\,\mathrm{Mpc}$
  (i.e., comparable to the 5NN distance 
  in the highest densities), 
   the reconstruction
  would have been  highly affected by shot noise  
  in the VIPERS regions with medium/low density.
  In those regions, the number of
  tracers inside a filter with small fixed aperture
  is very low: considering e.g.~that at $z\simeq0.7$ the mean surface density of
  tracers is about 85 objects per $\mathrm{deg}^2$,  
  only three tracers are expected on average
  within a cylinder having 
  $R\simeq3\,h_{70}^{-1}\,\mathrm{Mpc}$.

  The use of cylinders, instead of e.g.~spherical filters, 
  minimises the impact of redshift-space 
  distortions \citep{Cooper2005}. 
  The depth along the line of sight ($2\,000\,\mathrm{km\,s}^{-1}$)  
  is optimal not only for spectroscopic redshifts, but also for 
  photometric ones after multiplying their PDF by $N(z)$ 
  as  described above. 
  The reconstruction  of the density field through the procedure 
  described here
  is extensively tested in \citet{Cucciati2014a},
  but using spherical filters with 
  $R_\mathrm{fixed}=7.1$ and $11.4\,h_{70}^{-1}\,\mathrm{Mpc}$
  (5 and 8\,Mpc if 
  $H_0=100\,\mathrm{km\,s^{-1}\,Mpc^{-1}}$).
  We verified 
  that the outcomes do not change 
  when replacing spheres with cylinders (Cucciati et al.~in prep.). 
  For a detailed comparison among different filters 
  (spheres or cylinders, fixed or adaptive apertures, etc.) 
  we refer to \citet{Kovac2010a} and \citet{Muldrew2012}.
  
  The local density contrast of a given galaxy
  is   
  \begin{equation}
  \delta(r_g,z_g,R_{\mathrm{5NN}})= 
  \frac{\rho (r_g,z_g,R_{\mathrm{5NN}}) - 
  \bar{\rho}(z_g) }
  {\bar{\rho}(z_g) },
  \label{delta}
  \end{equation}  
  where we estimate $\bar{\rho}(z_g)$ 
  as a function of
  redshift by smoothing the spectroscopic
  distribution $N(z)$ with the $V_\mathrm{max}$ statistical
  approach,  in a similar way to \citet{Kovac2010a}. 
  For galaxies near the survey edges we correct $\delta$
  as done in \citet{Cucciati2006}, i.e.~by rescaling the measured
  density by the fraction of the cylinder volume within the
  survey borders. We notice however that the scatter 
  in the density field reconstruction is mainly due
  to the survey strategy (e.g.,~the sampling rate).  
  The impact of border effects is much smaller, 
  and becomes significant only when  most of
  the cylinder volume  ($>50\%$) is outside the survey area. 
  When it happens,  
  we prefer to discard the object from the sample. 
  We also remove galaxies for which  the cylinder 
  is inside the survey borders, but  less than 60\% is included  
  in the spectroscopically
  observed volume 
  (e.g., when more than 40\% of it 
  falls in gaps or in a missing VIMOS quadrant).
  In that case the density contrast should rely mostly on photometric neighbours, 
  and our estimate would be less accurate.
  With these two constraints, we excluded about 9\% of
  the objects (almost all located on the edges of the survey).



  \section{Environment and galaxy type classification}
  \label{Classifications}
  
  A fundamental step in this work is to identify the
  galaxies residing in two opposite environments, 
  i.e.~regions 
  of low density (LD) and high density (HD).
  Broadly speaking, 
  the former ones are regions 
  without a pervasive presence of 
  cosmic structure,
  whereas the latter are 
  associated with the highest peaks of the matter distribution. 
  Ideally, one would like to link these definitions to the total matter density. 
  However, since the dark matter component is not directly
  observed, any  classification has to rely on 
  some proxy of the overall density field. Our classification 
  relies on the galaxy density contrast evaluated in Sect.~\ref{Density contrast}.
  In addition to this, we  divide galaxies by type, 
  to work in each environment with
  active and passive objects separately.
  
  \subsection{Galaxy environments}
  \label{Environment definition}
  
  \begin{figure}
  \centering
  \includegraphics[width=0.5\textwidth]{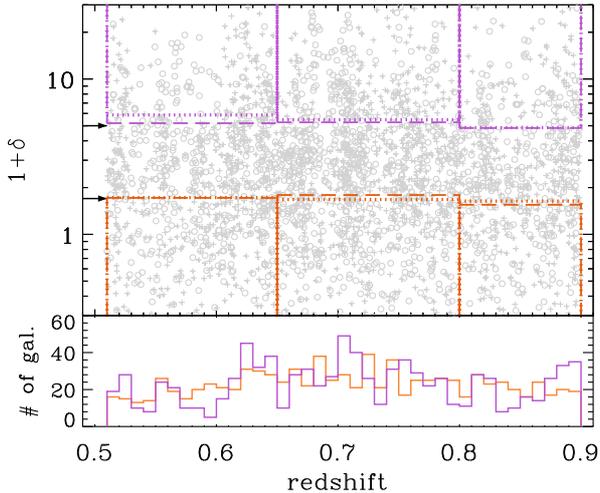} 
  \caption{\textit{Upper panel:} galaxy density contrast 
  of a mass-limited sample having 
  $\log(\mathcal{M/M_\odot})>$10.86. 
  Galaxies from the W1 field are marked with plus signs, 
  from W4 with open circles.
  For each redshift bin, galaxies below the 25th (above the 75th)
  percentile of the $1+\delta$ distribution 
  are enclosed by orange (violet) rectangles    
  (dotted lines for W1, dashed lines for W4). 
  The two tresholds that define LD and HD, 
  as discussed in Sect.~\ref{Environment definition}, 
  are marked by an arrow on the left side of the plot.
  \textit{Lower panel:}  combining the two fields together,
  histograms represent the redshift distribution 
  of the LD and HD sub-sample, in orange and violet respectively.  
  }
  \label{d1d4lim}
  \end{figure}

  In our analysis, we discriminate 
  LD from HD environments 
  by means of the 
  local density contrast. 
  We include in the LD (HD) sample galaxies
  that have a density contrast smaller (larger) than
  a certain value of  $\delta$. These  thresholds
  can be fixed according to some physical prescription 
  \citep[e.g.~to match detections of galaxy groups or clusters, as in][]{Kovac2010a},
  or determined in a relative way, e.g.~by considering the extreme tails 
  of the $1+\delta$ distribution.
  Following the latter approach, 
  \citet[][zCOSMOS 10k sample]{Bolzonella2010}
  assume as reference for low and high densities
  the 25th and 75th percentile (i.e., first and third quartile)
  of the $1+\delta$ distribution,
  respectively. The authors compute the distribution 
  in each of their redshift bins, independently;
  however,  we notice that  the thresholds  they
  estimate in bins between  $z=1$ and 0.5 are almost constant
  \citep[see also][Fig.~9]{Peng2010}.

  Similarly to \citet{Bolzonella2010},
  we compute the distribution of $1+\delta$
  (distinctly in W1 and W4)
  within three redshift bins: 
  $0.51<z\leqslant0.65$, $0.65<z\leqslant0.8$, $0.8<z\leqslant0.9$. 
  We choose this partition to probe a volume sufficiently large 
  in each bin  ($\gtrsim7\times10^{6}\,h_{70}^{-3}\,\mathrm{Mpc}^{3}$). 
  Moreover, the resulting median redshifts 
  ($\langle z \rangle$, see Table~\ref{tabmlim}) 
  correspond to nearly equally-spaced time steps (0.6--0.7\,Gyr). 
  We will adopt the same redshift bins 
  to estimate the
  mass functions in Sect.~\ref{Gsmf results}.
  Here we take into account only galaxies 
  with $\log(\mathcal{M/M_\odot})>10.86$,
  to work with a complete sample 
  in all the $z$-bins. Indeed, 
  such a value corresponds to the stellar mass 
  limit of the passive population at $z\simeq0.9$ 
  (see Sect.~\ref{Gsmf estimate} and 
   Table \ref{tabmlim}). 
  The resulting 25th and 75th percentiles 
  ($\delta_\mathrm{LD}$ and $\delta_\mathrm{HD}$) 
  vary among the three $z$-bins and the two fields 
  by less than $\sim20\%$, 
  namely
  $\delta_\mathrm{LD}$ assumes values 
  between 0.55 and 
  0.79, while   
    $3.84<\delta_\mathrm{HD}<5.87$.    
  These changes do not represent a monotonous
  increase  as a function of redshift, 
  but rather
  random variations between one $z$-bin and another,
  and between one field and the other (see Fig.~\ref{d1d4lim}). 
  
  In Appendix \ref{Appendix} we confirm, by means of 
  cosmological simulations, that  
  the small fluctuations of the percentile thresholds do 
  not reflect an evolution in $z$. In fact, they 
  are mainly due to sample variance, and 
  do not reflect the growth of structure over cosmic time. 
  Moreover, we verified that there is no bias introduced by the VIPERS 
  selection effect. 
  Therefore,  we can safely use constant 
  thresholds to classify LD and HD environments in VIPERS: 
  we consider  
  galaxies with $1+\delta<1.7$ 
  as belonging to LD, and 
  galaxies with $1+\delta>5$ to  HD. 
  These limits, applied from $z=0.9$ to 0.51,
  are the mean of 25th and 75th 
   percentiles computed above 
  (see Fig.~\ref{d1d4lim}).  
  Despite the name we chose for sake of clarity, 
  we note that the HD regions in VIPERS 
  have actually intermediate
  densities in absolute terms. 
  Very concentrated structures, such as massive galaxy clusters, 
  typically have $1+\delta\simeq15$--20 \citep{Kovac2010a} 
  and should approximately match the upper 5\%  of environmental
  density. However the HD environment we defined, 
  although on average less extreme, 
  is certainly  interesting to study, since it has evolved more recently 
  than typical clusters \citep[][]{Smith2005,Fritz2005}.
  
  As stated above, with the 5NN we tend to probe 3--8$\,h_{70}^{-1}\,
  \mathrm{Mpc}$. 
  Hence, if a certain environmental mechanism were efficient 
  at smaller scale, its trace could be ``diluted'',  
  or even vanish, in our analysis. 
  However, this is not the case, as we will show in the following. 
  Environmental dependencies at large scales  
  have already been measured e.g.~in 
  \citet{Cucciati2006} \citep[see also][]{Bassett2013,Hearin2015}. 
  These findings 
  can be due to physical mechanisms operating at distances larger 
  than the halo virial radius
  \citep[e.g.][]{LuT2012}. Another possibility is that a connection 
  between large-scale environment and halo properties 
  preserves the small-scale signal 
  even when working with lower resolutions.   
  Supporting the latter argument, 
  \citet{Haas2012} demonstrated  that 
  estimators based on a number of neighbours $2\leqslant N \leqslant10$
  correlate equally well with host halo mass.

  Further details about the estimate of the density field 
  are given in Appendix \ref{Appendix}. Among the tests 
  described there,  
  we also evaluate the 
  purity and completeness of our LD and HD samples. 
  By working on mock galaxy catalogues, we show that 
  our method is not hindered 
  by the  VIPERS selection function:  more than 
  70\%  of LD/HD galaxies are expected to be assigned   
  to the correct environment, while a small tail 
  of objects  ($<8\%$) end up in the opposite one 
  as interlopers.

  \subsection{Classification of galaxy types}  
   \label{Galaxy type classification}  
  \begin{figure}
  \includegraphics[width=0.99\columnwidth]{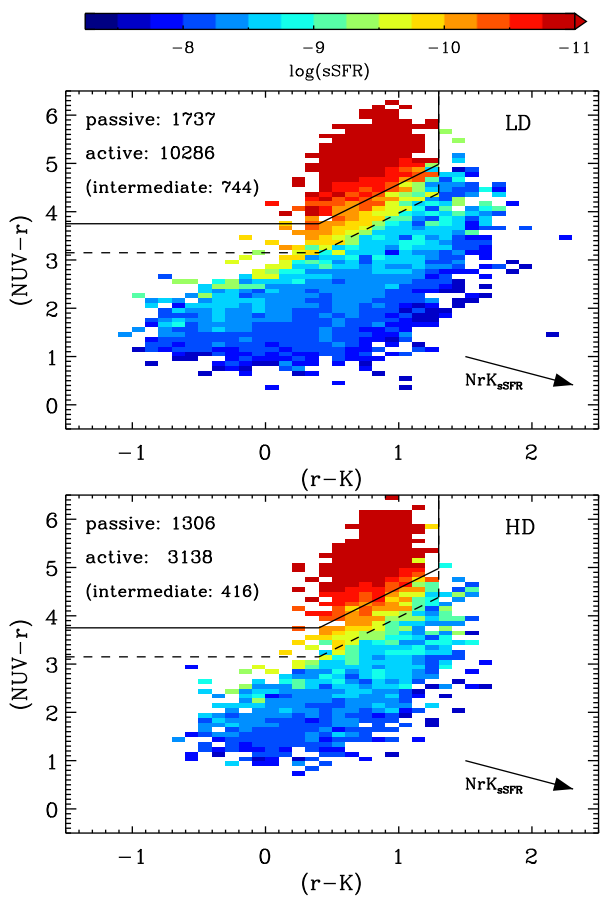} 
  \caption{The $\mathrm{NUV}rK$ diagram of the VIPERS galaxies 
  between $z=0.51$ and 0.9, in the LD 
  environment (top panel) and in HD (bottom panel).
   According to our classification, 
   passive galaxies lie above the solid line (defined in Eq.~\ref{nrkcut}), 
   while the dashed line (Eq.~\ref{nrkcut2}) divides galaxies with
   intense star formation (bottom part of the diagram)
   from those having low-sSFR (see text). 
   In the remainder of the paper, the two classes will be treat as 
   a whole  sample of active objects. 
   Their number (and the number of passive galaxies) in each environment 
   is shown in the top-left corner of the plots.
   Each colour-coded pixel represents the median sSFR 
   of the galaxies inside it, 
   estimated by means of SED fitting.
   \citet{Arnouts2013} find that in this diagram the sSFR increases 
   as moving along the direction 
   $[(r-K),(\mathrm{NUV}-r)]=[(r-K)_0+\sin(54\degr),
   (\mathrm{NUV}-r)_0-\cos(54\degr)]$, 
    identified by the bottom-right vector $NrK_\mathrm{sSFR}$
    (note that the different scale in $x$- and $y$-axis warps the angles).
   }
  \label{nrkdiag}
  \end{figure}
  
  In order to separate active and passive galaxies, we apply  the
  method described in \citet{Arnouts2013}, based on the 
  $(\mathrm{NUV}-r)$ vs $(r-K)$ diagram ($\mathrm{NUV}rK$ in the following). 
  With this method, recent star formation
  on a scale of $10^6$--$10^8$\,yr
  is traced by the $(\mathrm{NUV}-r)$ colour \citep[][]{Salim2005},
  while  $(r-K)$ is sensitive to 
  the inter-stellar medium (ISM) absorption. 
  The absolute magnitudes used here 
  have been estimated as detailed  in Sect.~\ref{Sed fitting},
  through the filters $\mathrm{NUV}$, $r$, and $Ks$
  of GALEX, MegaCam, and WIRCam respectively.  
  It should be noticed that our redshift range ($0.51<z\leqslant0.9$) is  
  fully within the interval $0.2 < z < 1.3$  used by 
  \citeauthor{Arnouts2013} in their analysis.
  Their diagram is similar to the $(U-V)$ vs $(V-J)$ plane proposed by 
  \citet{Williams2009}, but  by sampling more extreme wavelengths
  it results in a sharper separation between quiescent and
  star-forming galaxies \citep[cf~also][]{Ilbert2013}.
  Moreover, 
  the position of  a galaxy in the $\mathrm{NUV}rK$ plane 
  correlates well with its infrared excess (IRX, i.e.~the 
  $L_\mathrm{IR}/L_\mathrm{NUV}$ ratio) 
  and specific SFR   
  ($\mathrm{sSFR}\equiv\mathrm{SFR}/\mathcal{M}_\odot$), 
  at least when $\log(\mathcal{M/M_\odot})\geqslant 9.3$
  \citep[for further details, see][]{Arnouts2013}.
  It should also be emphasised that 
  with classification methods based on a single-colour bimodality,
  the passive sample is partially contaminated by star-forming 
  galaxies reddened by dust, as shown e.g.~by \citet{Moresco2013}.
  With the $\mathrm{NUV}rK$, the simultaneous use of two colours disentangles
  those different populations.

  As illustrated in Fig.~\ref{nrkdiag} (solid line), we regard a galaxy 
  as passive when
  \begin{equation}
   \begin{array}{l}
  (\mathrm{NUV} - r) > 3.75  \quad \mathrm{and}\\ 
     (\mathrm{NUV} - r) > 1.37 (r - K) + 3.2  \quad \mathrm{and}\\ 
   (r - K) < 1.3 \, .
  \end{array}
  \label{nrkcut}
  \end{equation}
  Active galaxies are located in 
  the complementary region of the diagram (i.e.~below the solid line 
  in Fig.~\ref{nrkdiag}).
   
  With respect to the definition of  \citet{Arnouts2013}
  we added a further cut, namely $(r-K)<1.3$.
  In this way we take  into account  the
  geometrical effect they observe after including the
  dust prescription of  \citet{Chevallard2013} in their
  analysis. According to that study, the reddest $(r-K)$ colours
  can be reached only by edge-on disc galaxies with
  a flat attenuation curve. 
  We also verified through a set
  of BC03 templates  that passive
  galaxies ($\mathrm{age}/\tau>4$) 
  have $(r-K)<1.15$.
  This result, considering the typical 
  uncertainties in magnitude estimates, 
  justifies the third
  condition in Eq.~(\ref{nrkcut}). With a similar argument,
  \citet{Whitaker2011} modify the passive \textit{locus}
  of \citet{Williams2009} diagram.
   
  In the $\mathrm{NUV}rK$,  sSFR increases as galaxies move along a
  preferential direction, 
  identified in Fig.\ref{nrkdiag} by the vector $NrK_\mathrm{sSFR}$. 
  Therefore, lines orthogonal to that direction 
  work as a cut in sSFR: for instance, 
  the diagonal boundary we defined for the passive \textit{locus} 
  roughly corresponds to 
  $\mathrm{sSFR}<10^{-11}\,\mathrm{yr}^{-1}$.
  We prefer to use $\mathrm{NUV}rK$ instead of selecting
  directly through the sSFR distribution, 
  since the SED fitting estimates of SFR are
  generally less reliable than colours \citep[][]{Conroy2009b},
  especially when far-IR data are not available.
  Nevertheless, it is worth noticing that the sSFR values 
  we obtained from \textit{Hyperzmass} are on average in good
  agreement with the $\mathrm{NUV}rK$ classification, 
  providing an additional confirmation  
  of its robustness (see Fig.~\ref{nrkdiag}).
  Among the galaxies we have classified as $\mathrm{NUV}rK$-passive, 
  about 95\%  have a (SED fitting derived) sSFR lower than
  $10^{-11}\,\mathrm{yr}^{-1}$  \citep[which is the typical cut used 
  e.g.~in][]{Pozzetti2010}.

  We also tested another boundary in the colour-colour space 
  (the dashed line in Fig.~\ref{nrkdiag}), namely 
  \begin{equation}    
  \begin{array}{l}
  (\mathrm{NUV} - r) > 3.15 \quad \mathrm{and} \\
  (\mathrm{NUV}- r) > 1.37 (r - K) + 2.6 \quad \mathrm{and} \\
  (r - K) < 1.3 \, .
  \end{array} 
  \label{nrkcut2}
  \end{equation}
  In this way we can delimit a region in the 
  $\mathrm{NUV}rK$ plane likely corresponding to the ``green valley'': 
  galaxies in between Equations  (\ref{nrkcut}) and (\ref{nrkcut2}) 
  are probably shutting off their star formation, 
  having  
  $\mathrm{sSFR} \simeq 10^{-10}\,\mathrm{yr}^{-1}$ 
  according to their SED fitting estimates 
  \citep[Fig.~\ref{nrkdiag}, but see also][]{Arnouts2013}.  
  We include these galaxies in the active sample, although 
  they are expected to be in transition towards the passive 
  \textit{locus}. We verify that removing them from the 
  active sample do not modify our conclusions. The typical features of this 
  ``intermediate'' galaxies will be explored in a future work.



  \section{Stellar mass functions in different environments}
  \label{Gsmf results}

  We now derive the stellar mass function of VIPERS galaxies within the 
  environments described in Sect.~\ref{Environment definition}, 
   also  separating active and passive subsamples. 
  The chosen  redshift bins are those  already adopted there (also reported in Table \ref{tabmlim}).  
  We describe our results and compare them  to what has been found by previous 
  surveys.

  \subsection{Methods}
  \label{Gsmf estimate}
  
  First, we determine the threshold $\mathcal{M}_\mathrm{lim}$
  above which our data can be considered complete in stellar mass.
  As explained below, $\mathcal{M}_\mathrm{lim}$ 
  depends on the flux limit of VIPERS ($i_\mathrm{lim}$), redshift,
  and galaxy type.
  Such a limit excludes stellar mass bins with large fractions of undetected objects.

  The estimate of $\mathcal{M}_\mathrm{lim}$ is complicated 
  by the wide range of $\mathcal{M}/L$. 
  To estimate such a limit    
  we apply the technique of \citet{Pozzetti2010}, which 
  takes into account typical $\mathcal{M}/L$
  of the faintest observed galaxies 
  (see also the discussion in D13, Sect.~3.1).   
  We keep separated
  the active sample from the passive one, 
  since  $\mathcal{M}/L$ depends on galaxy type.
  For each population we select 
  the 20\% faintest objects 
  inside each redshift bin. 
  We rescale their stellar masses at the
  limiting magnitude: 
  $\log(\mathcal{M}(i\!=\!i_\mathrm{lim})/\mathcal{M_\odot}) \equiv
  \log(\mathcal{M}/\mathcal{M_\odot})+0.4(i-i_\mathrm{lim})$.
  For the active and passive sample respectively,
  we define $\mathcal{M}_\mathrm{lim}^\mathrm{act}$ and
  $\mathcal{M}_\mathrm{lim}^\mathrm{pass}$ 
  to be equal to the 98th and 90th percentile of the corresponding 
  $\mathcal{M}(i\!=\!i_\mathrm{lim})$ distributions.
  We choose a higher percentile level for active galaxies 
  to take into account the larger scatter they have in $\mathcal{M}/L$.  
  Results are reported in Table \ref{tabmlim}.
  The increase of the limiting mass toward higher redshifts 
  is due to dimming,
  while $\mathcal{M}_\mathrm{lim}^\mathrm{act}$ is always smaller
  than $\mathcal{M}_\mathrm{lim}^\mathrm{pass}$
  because passive SEDs have on average larger $\mathcal{M}/L$.
  For the total GSMF we will use 
  $\mathcal{M}_\mathrm{lim}^\mathrm{pass}$
  as a conservative threshold; 
  a direct estimate by applying the technique of \citet{Pozzetti2010} 
  to the whole sample would result in lower
  values by about 0.2--0.3 dex, 
  because of the mixing of galaxy types 
  \citep[cf D13;][]{Moustakas2013}.
  
  We then  estimate the stellar mass functions
  by means of two methods, namely the $1/V_\mathrm{max}$ method
  \citep{Schmidt1968} and the one devised by 
  \citet[][hereafter STY]{Sandage1979}.
  The former is non-parametric, whereas the latter
  assumes the GSMF to be modelled by the
  \citet[][]{Schechter1976} function
  \begin{equation}
  \Phi(\mathcal{M})\mathrm{d}\mathcal{M} =
  \Phi_\star \left( \frac{\mathcal{M}}{\mathcal{M}_\star} \right)^{\alpha} 
  \exp\left( - \frac{\mathcal{M}}{\mathcal{M}_\star} \right) 
  \frac{\mathrm{d}\mathcal{M}}{\mathcal{M}_\star}  \: .
  \label{schfun}
  \end{equation}
  Both of them are
  implemented in the software package ALF \citep{Ilbert2005}.  
  
  The $1/V_\mathrm{max}$ method 
  gives the comoving galaxy density  
  in a certain stellar mass bin (e.g., between $\mathcal{M}$ 
  and $\mathcal{M}+\mathrm{d}\mathcal{M}$):    
  \begin{equation}
  \Phi(\mathcal{M})\mathrm{d}\mathcal{M}= 
  \sum_{j=1}^N \frac{w_j}{V_{\mathrm{max}\,j}} \;,
  \end{equation}    
  where $V_{\mathrm{max}}$ is the comoving 
  volumes in which a galaxy (out of the $N$ detected in the 
  given bin) would be observable,   and  $w$ is 
  the statistical weight described in Sect.~\ref{Dataspec}. 
  Usually, to measure   $V_\mathrm{max}$ 
   one needs to know   
  the sky coverage of the survey, and 
  the minimum and maximum redshifts 
  at which the object drops out the magnitude range of detection. 
  However, 
  considering the whole surveyed area is not  formally correct 
  when dealing with HD/LD galaxies -- as well as galaxies in 
  clusters or groups --  because those objects have no chance 
  (by definition) to be observed outside their environment. 
  In other words, we need to  reconstruct   
  the comoving volumes occupied by the HD/LD regions 
  and take them into 
  account,  instead of  
  the total VIPERS volume, to estimate the $V_\mathrm{max}$ values. 
  This new approach  is described in detail in Appendix \ref{Voronoi}.  
  It allows us to properly normalise the stellar mass functions 
  in Fig.~\ref{mfcfvmax}, 
   In the same Appendix we also describe how we estimated the 
   uncertainty due to cosmic variance, by means of galaxy mocks.    
    We include this uncertainty  
    in the error budget of the total GSMFs, along with Poisson noise, 
    while for the active and passive samples  
    we compute  errors assuming Poisson statistics  only.  
    In plotting each GSMF, the $1/V_\mathrm{max}$ points are centred at 
    the median stellar mass of their bin. 
    We evaluate the error on this position,   
    i.e.~the error bar on the $x$-axis, 
     by considering 100 simulated Monte Carlo samples 
     in which the uncertainty of $\log(\mathcal{M/M_\odot})$ 
     is randomly assigned from a 
     Gaussian 0.2\,dex large. After binning those samples, 
     the median stellar mass 
     within each bin shows a variance on average 
     smaller than 0.05\,dex, fully negligible in the treatment.

\begin{table}[h]
\caption{Stellar mass completeness: thresholds 
 for active and passive galaxies in the redshift bins adopted in this work. 
 These limits are valid both in LD and HD regions;
 $\mathcal{M}_\mathrm{lim}^\mathrm{pass}$ is used also for the whole
 galaxy sample. In addition, the median redshift of each bin is 
 reported in the second column.}
\label{tabmlim}
\setlength{\extrarowheight}{1ex}
\centering
\begin{tabular}{lccc}
\hline \hline
 redshift range   & $\langle z \rangle$    
 & $\log(\mathcal{M_\mathrm{lim}^\mathrm{act}}/\mathcal{M_\odot})$ & 
 $\log(\mathcal{M_\mathrm{lim}^\mathrm{pass}}/\mathcal{M_\odot})$ \\ [1ex] 
\hline
$0.51<z \leqslant0.65$  & 0.60  & 10.18   & 10.39      \\
$0.65<z \leqslant0.8$    & 0.72  & 10.47   & 10.65   \\
$0.8<z \leqslant0.9$      & 0.84  & 10.66   & 10.86      \\ [1ex]
\hline
\end{tabular}
\end{table}

  The STY method determines the 
  parameters $\alpha$ and $\mathcal{M}_\star$
  of Eq.~(\ref{schfun})
  through a maximum-likelihood approach. 
  The associated uncertainties come from  the confidence
  ellipsoid at 1$\sigma$ level. 
  In the highest redshift bin, i.e.~$0.8<z<0.9$,  we are limited
  to $\log\mathcal{M/M_\odot}\gtrsim10.7$ and therefore 
  we prefer to keep $\alpha$ fixed to the value found in the
  previous $z$-bin.  
  The third parameter ($\Phi_\star$) is computed independently, 
  to recover the galaxy number density after integrating the 
  Schechter function \citep[see][]{Efstathiou1988,Ilbert2005}. 
  Also in this case we consider the comoving volumes occupied by the 
  two environments (Appendix \ref{Voronoi}).

  The STY estimates, along with their  uncertainties, 
  are listed in Table \ref{tab_mstar}. 
  Complementary to the $1/V_\mathrm{max}$ estimator in 
  many aspects, this method is unbiased 
  with respect to density inhomogeneities
  \citep[see][]{Efstathiou1988}. 
  We verified that the $1/V_\mathrm{max}$ outcomes are 
  reliable by comparing its outcomes not only with the STY but
  also with another non-parametric estimator \citep[i.e., the
  stepwise maximum-likelihood method of][]{Efstathiou1988}. 
  These multiple estimates strengthen our results, as  
  the different methods turn out to be in good agreement (Fig.~\ref{gsmf1}).
  In particular, this fact validates the completeness limits 
   we have chosen, because the estimators would diverge at 
  $\mathcal{M>M}_\mathrm{lim}$ 
  if some galaxy population were missing 
  \citep[see][]{Ilbert2004}.

\begin{table*}
\caption{GSMF in low- and high-density regions: Schechter parameters resulting from the STY method, when applied to different galaxy populations at different redshifts. Before fitting data at $0.8<z<0.9$, $\alpha$ has been fixed to the value of the previous $z$-bin.}
\label{tab_mstar}
\setlength{\extrarowheight}{1ex}
\centering
\begin{tabular}{lcccccc}
\hline\hline
galaxy sample & $\alpha$ & $\log\mathcal{M}_\star$  & $\Phi_\star$  & $\alpha$ & $\log\mathcal{M}_\star$ & $\Phi_\star$\\
 & & $[h_{70}^{-2}\,\mathcal{M}_\odot]$ & $[10^{-3}\,h_{70}^3\,\mathrm{Mpc}^{-3}]$ & & $[h_{70}^{-2}\,\mathcal{M}_\odot]$ & $[10^{-3}\,h_{70}^3\,\mathrm{Mpc}^{-3}]$ \\ 
[1ex] \hline
$ 0.51<z<0.65 $ & \multicolumn{3}{c}{low density } & \multicolumn{3}{c}{high density} \\
     total & $-0.95^{   + 0.16}_{   -0.16}$ & $10.77^{   + 0.06}_{   -0.05}$ & $ 1.27^{   + 0.17}_{   -0.19}$ & $-0.76^{   + 0.14}_{   -0.13}$ & $11.01^{   + 0.06}_{   -0.06}$ & $ 4.60^{   + 0.59}_{   -0.63} $ \\
   passive & $-0.49^{   + 0.20}_{   -0.20}$ & $10.76^{   + 0.06}_{   -0.06}$ & $ 0.73^{   + 0.06}_{   -0.08}$ & $-0.00^{   + 0.18}_{   -0.18}$ & $10.89^{   + 0.06}_{   -0.05}$ & $ 3.51^{   + 0.16}_{   -0.16} $ \\
    active & $-0.87^{   + 0.20}_{   -0.19}$ & $10.51^{   + 0.06}_{   -0.06}$ & $ 1.18^{   + 0.16}_{   -0.19}$ & $-0.93^{   + 0.19}_{   -0.18}$ & $10.77^{   + 0.08}_{   -0.08}$ & $ 2.71^{   + 0.55}_{   -0.57} $ \\
[1ex] \hline
$ 0.65<z<0.80 $ & \multicolumn{3}{c}{low density } & \multicolumn{3}{c}{high density} \\
     total & $-0.52^{   + 0.32}_{   -0.31}$ & $10.72^{   + 0.07}_{   -0.06}$ & $ 1.14^{   + 0.07}_{   -0.11}$ & $-0.80^{   + 0.23}_{   -0.22}$ & $10.99^{   + 0.08}_{   -0.07}$ & $ 3.83^{   + 0.55}_{   -0.69} $ \\
   passive & $-0.14^{   + 0.40}_{   -0.39}$ & $10.73^{   + 0.09}_{   -0.08}$ & $ 0.51^{   + 0.03}_{   -0.04}$ & $-0.40^{   + 0.28}_{   -0.27}$ & $10.97^{   + 0.09}_{   -0.07}$ & $ 2.42^{   + 0.18}_{   -0.32} $ \\
    active & $-1.26^{   + 0.32}_{   -0.31}$ & $10.69^{   + 0.10}_{   -0.09}$ & $ 0.79^{   + 0.20}_{   -0.24}$ & $-0.91^{   + 0.31}_{   -0.30}$ & $10.78^{   + 0.10}_{   -0.09}$ & $ 2.54^{   + 0.51}_{   -0.65} $ \\
[1ex] \hline
$ 0.80<z<0.90 $ & \multicolumn{3}{c}{low density } & \multicolumn{3}{c}{high density} \\
     total & $-0.52 $ & $10.64  ^{ + 0.05  }_{ -0.04}$ & $ 1.16^{   + 0.08}_{   -0.08}$ & $-0.80 $ & $10.85  ^{ + 0.05  }_{ -0.04}$ & $ 4.59^{   + 0.33}_{   -0.33} $ \\
   passive & $-0.14 $ & $10.66  ^{ + 0.06  }_{ -0.05}$ & $ 0.36^{   + 0.04}_{   -0.04}$ & $-0.40 $ & $10.88  ^{ + 0.06  }_{ -0.05}$ & $ 1.76^{   + 0.18}_{   -0.18} $ \\
    active & $-1.26 $ & $10.70  ^{ + 0.07  }_{ -0.07}$ & $ 0.85^{   + 0.05}_{   -0.05}$ & $-0.91 $ & $10.75  ^{ + 0.07  }_{ -0.06}$ & $ 3.35^{   + 0.25}_{   -0.25} $ \\
\hline
\end{tabular} 
\end{table*}

  \subsection{Results}
  \label{Results}

    \begin{figure*}
  \includegraphics[width=0.9\textwidth]{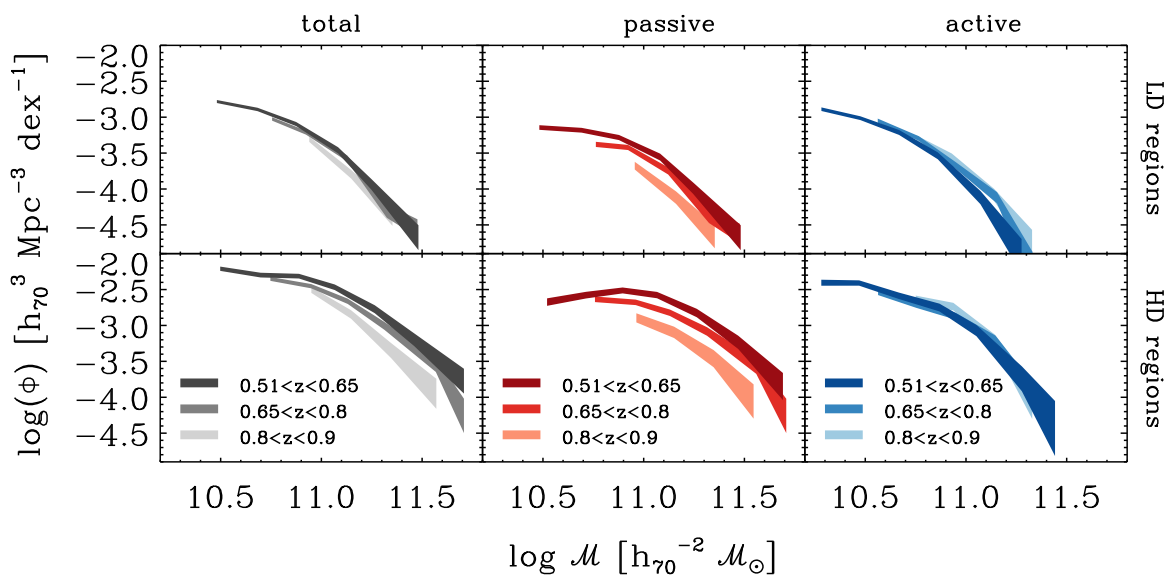}
  \caption{Evolution of the GSMF in the different 
  VIPERS environments. Total, passive, and active samples are in 
  black, red, and blue colours respectively. Each shaded area is obtained 
  from the $1/V_\mathrm{max}$ estimates  adding the corresponding Poissonian 
  uncertainty (see Sect.~\ref{Gsmf estimate} and 
  Appendix \ref{Voronoi} for details); only estimates above the 
  stellar mass completeness limit are considered. }
  \label{mfcfvmax}
  \end{figure*}

  \begin{figure*}
  \includegraphics[width=0.99\textwidth]{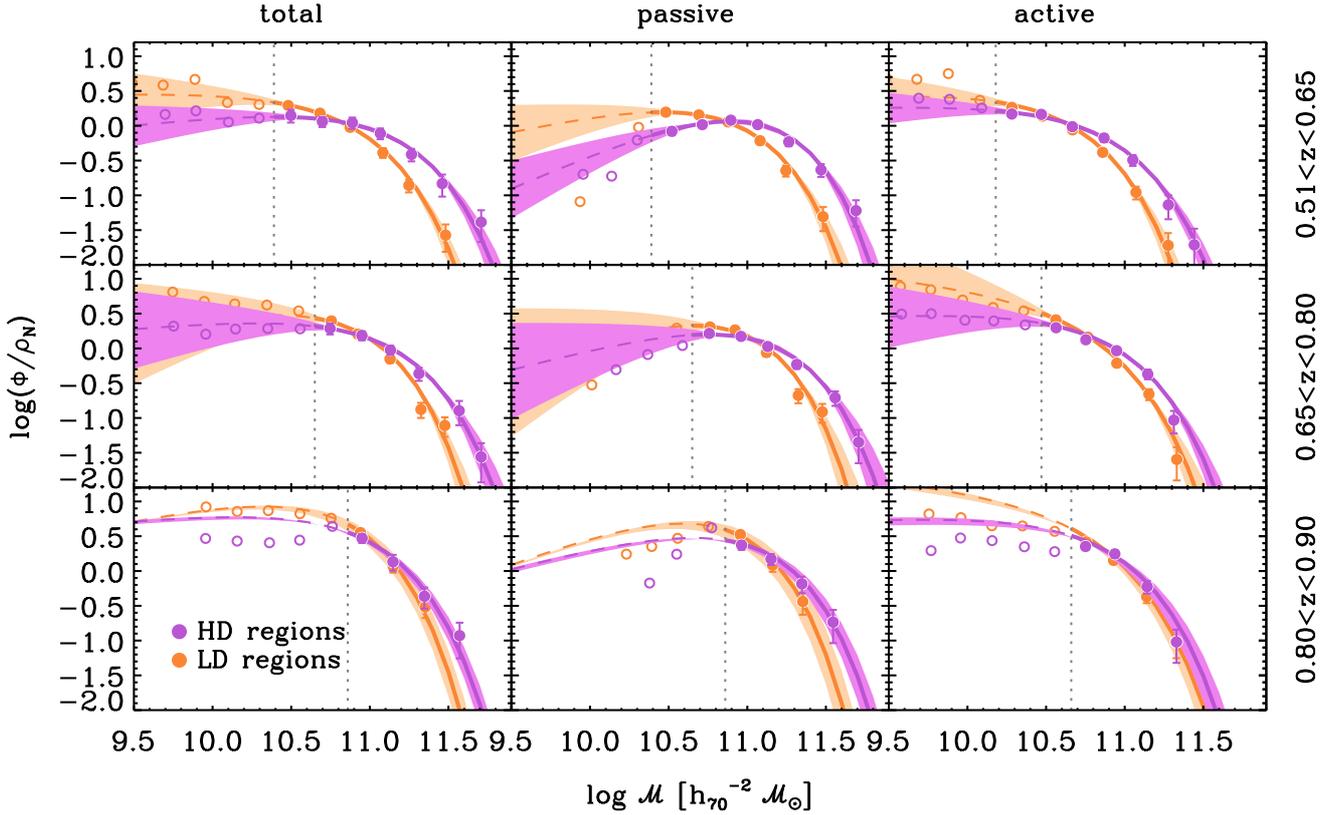}
  \caption{Stellar mass functions of galaxies at low density (orange symbols) 
  and high density (violet symbols) in three different redshift bins, namely
  $0.51<z\leqslant0.65$, $0.65<z\leqslant0.8$, and $0.8<z\leqslant0.9$.  
  Right-side panels show the GSMFs
  of active galaxies, while central panels refer to passive ones. The GSMFs
  of the whole sample in the same $z$-bins are shown on the left.  
  In each plot, filled (open) circles represent the $1/V_\mathrm{max}$ 
  points above (below) the completeness mass $\mathcal{M}_\mathrm{lim}$
  (vertical dot line),
  with error bars (shown only above $\mathcal{M}_\mathrm{lim}$) 
  that accounts for Poisson uncertainty. 
  In the total GSMFs, also the uncertainty due to cosmic variance is 
  added in the error bars (note that in some cases the error bar is smaller 
  than the size of the points).
  Solid lines represent the Schechter functions estimated through the STY method,
  with the 1$\sigma$ uncertainty highlighted by a shaded area.
  With this estimator all the Schechter parameters are free, except at
  $0.8<z\leqslant0.9$, where $\alpha$ is fixed to the value found in the
  previous $z$-bin (see Table \ref{tab_mstar}).
  To compare the shape of mass functions in LD and HD, we renormalise them
  in such a way that their number density ($\rho_N$) is equal to unity when we
  integrate the GSMF at $\mathcal{M}>\mathcal{M}_\mathrm{lim}$.
  }
  \label{gsmf1}
  \end{figure*}  
  

\begin{figure}
\includegraphics[width=0.99\columnwidth]{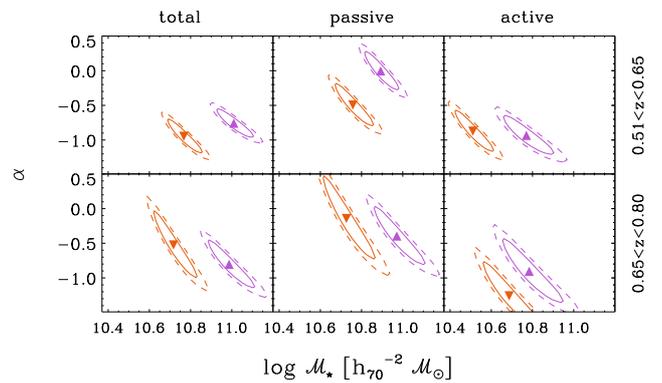}
\caption{\citet{Schechter1976} parameters (filled symbols) 
 of the GSMFs at redhisft 
 $0.51<z<0.65$ and $0.65<z<0.8$, where $\alpha$ was let free during 
 the STY fitting (cf Fig.~\ref{gsmf1}). 
 The solid- and dashed-line contours represent 
 respectively the 68.3 and 90\%  CL. Orange lines and downward triangles 
 are the estimates for galaxies in the LD regions, violet lines and 
 upward triangles are used for the HD ones.  
 Each panel concerns a different sample (total, passive, and active 
 galaxies from left to right). All the values are obtained by using the 
 algorithms contained in the ALF suite \citep{Ilbert2005}.  
}
\label{schellips}
\end{figure}
  
  The GSMFs computed in this Section are shown
  in Fig.~\ref{mfcfvmax} and \ref{gsmf1}.  
  In the former, to show their evolution, 
  we superimpose the mass functions 
  at  different redshifts, namely 
  $0.51<z\leqslant0.65$, $0.65<z\leqslant0.8$, 
  and $0.8<z\leqslant0.9$ (median redshift $\tilde{z}\sim 0.6$, 0.72, 0.84). 
  On the other hand, in Fig.~\ref{gsmf1},  we renormalise the GSMFs
  in such a way that their number density is equal to unity when we
  integrate the GSMF at $\mathcal{M}>\mathcal{M}_\mathrm{lim}$.  
  With this kind of rescaling we can directly 
  compare the shape that the GSMF has in the two environments. 
  In both Figures, the mass functions of different galaxy types  
  (total, passive, and active samples) are plotted in distinct columns.

  Our results are particularly intriguing  in the high-mass regime, where VIPERS
  benefits from its large number statistics. 
  Figure \ref{mfcfvmax} shows a different growth of stellar mass in LD and HD 
  environments. Regarding the total galaxy sample, there is  
  a mild increase of the HD high-mass tail over cosmic time 
  (bottom-left panel), an increase that 
  is not observed  neither in LD (top-left panel) 
  nor in the GSMF of the whole VIPERS 
  volume (D13).  This trend  seems to be due to the passive
  population (central panels) and will be investigated  in Sect.~\ref{Discussion}. 
  
  Also looking at the shape of the GSMFs, there is a remarkable difference 
  between LD and HD galaxies (Fig.~\ref{gsmf1}). 
  At $z\leqslant0.8$,  a large fraction of massive galaxies inhabits
  the densest regions, resulting in a higher exponential tail of
  the HD mass function with respect to the LD environment. 
  At higher redshifts this difference 
  becomes less evident.
  Quantitatively, 
  the difference is well described by the Schechter 
  parameter $\mathcal{M}_\star$, which is larger in
  the HD regions (see  the likelihood contours for $\alpha$ 
  and $\mathcal{M}_\star$ shown in Fig.~\ref{schellips}). 
  For the total sample, in the first and second
  redshift bin, $\Delta\mathcal{M}_\star\equiv 
  \log(\mathcal{M}_{\star,\mathrm{HD}}/\mathcal{M}_{\star,\mathrm{LD}})
  =0.24\pm0.12$ and $0.27\pm0.15\,\mathrm{dex}$ respectively. 
  A similar deviation appears at  $0.8<z\leqslant0.9$
  ($\Delta\mathcal{M}_\star=0.21\pm0.11\,\mathrm{dex}$)
  although in that case  
  the formal  $\mathcal{M}_\star$ uncertainty 
  has been reduced by keeping $\alpha$ fixed in the fit. 
  The behaviour seen for the whole sample is also signature of 
  the GSMFs divided by galaxy types (Fig.~\ref{gsmf1}, 
  middle and right panels).

  At intermediate masses, our analysis  becomes less stringent.
  Given the completeness limit of VIPERS,  
  it is difficult to constrain the power-law slope of the GSMF.
  We find that $\alpha_\mathrm{HD}$ and $\alpha_\mathrm{LD}$ are
  compatible within the errors, with the exception of the passive sample
  at $0.51<z\leqslant0.65$, for which 
  the stellar mass function is  steeper in the LD regions.

  \subsection{Comparison with previous work}
  \label{Gsmf comparison}
  
  The comparison with other authors is not always 
  straightforward, given the different definitions of 
  environment and  galaxy types. Besides that, also the selection function 
  (and the completeness) change from one survey to another. 
  A piece of work with an approach
  very similar to ours is \citet{Bolzonella2010}.
  In that paper,  low- and high-density 
  galaxies in the zCOSMOS survey ($0.1<z<1.0$) are classified
  by means of the galaxy density contrast (derived 
  from the 5NN, as in our case).\footnote{
  For sake of
  simplicity, we use our notation (LD and HD) also when referring to the 
  low-/high-density galaxies of \citet{Bolzonella2010}, which are 
  named D1 and D4 in the original paper.} 
   \citeauthor{Bolzonella2010} observe a higher fraction
  of massive galaxies in overdense regions, although within
  the uncertainties of the GSMF estimates. 
  Down to the redshift range not reached by VIPERS
  ($0.1<z<0.5$) they also  find
  an upturn of the high-density GSMF 
  below $\log \mathcal{M/M_\odot}\lesssim10$.

\begin{figure}  
  \includegraphics[width=0.99\columnwidth]{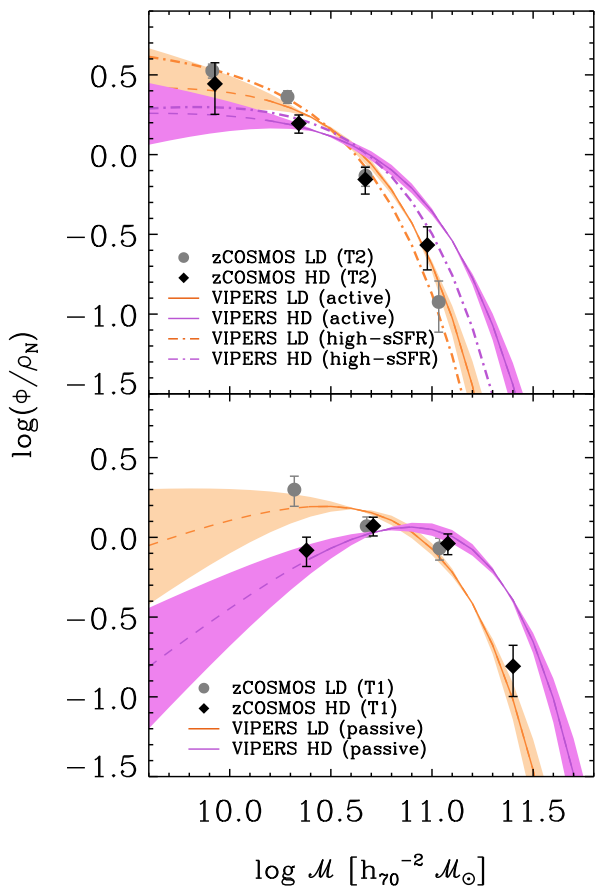}
  \caption{VIPERS (this work) and zCOSMOS  \citep{Bolzonella2010} 
   stellar mass functions of galaxies in LD/HD regions (orange/violet 
   and grey/black colours, 
   see the legend in the top-right corner of each panel).  
   The 
   comparison is restricted to a single 
   redshift bin that is similar in the two surveys 
   ($0.5<z<0.7$ in zCOSMOS, $0.51<z<0.65$ in VIPERS). 
   All the GSMFs are rescaled in order to have 
   $\rho_N(\mathcal{M}>\mathcal{M}_\mathrm{lim})=1$, 
   as in Fig.~\ref{gsmf1}.   
   In both panels, solid lines 
   represent the STY estimates for the various galaxy samples,   
   with a shaded area encompassing the 1$\sigma$ uncertainty 
   (the line is dashed below the stellar mass limit). 
   Filled circles and diamonds are the $1/V_\mathrm{max}$ 
   determinations of the  GSMFs of zCOSMOS 
   (LD and HD respectively). 
   The \textit{upper panel} includes the stellar mass functions 
   of star-forming galaxies,  classified  by \citet{Bolzonella2010} 
   according to their  photometric types (T2, i.e.~late-type galaxies), 
   and by means of the $\mathrm{NUV}rK$ diagram in our analysis. 
   We also show with dot-dashed lines  
   the stellar mass function of the VIPERS galaxies having high sSFR  
   (i.e., those remaining after removing the 
   $\mathrm{NUV}rK$-intermediate objects from the active sample). 
   In the \textit{lower panel}, the VIPERS passive  
   sample and the zCOSMOS early-type galaxies 
   (i.e., T1 spectrophotometric types) are considered. 
   }
   \label{zcos-vip}
  \end{figure}

  In Fig.~\ref{zcos-vip} we  directly compare our GSMFs 
  to those of 
  \citet{Bolzonella2010}  in a redshift bin 
  that is similar in the two analyses ($0.5<z<0.7$ in their paper, 
  $0.51<z<0.65$ in ours). We find a good agreement 
   for both passive and active   
  galaxies.\footnote{When considering 
  the next bin of \citeauthor{Bolzonella2010}, i.e.~$0.7<z<1$, 
  we also  found a fairly good agreement with 
  our data at $0.65<z<0.9$. 
  However we preferred to show the $z$-bin 
  where the stellar mass limit 
  is lower.} 
  With respect to the latter sample, 
  a better accordance  is reached  considering only high-sSFR 
  galaxies, i.e.~when we remove the ``intermediate'' 
  objects that lie 
  between the borders (\ref{nrkcut}) and (\ref{nrkcut2}) of 
  the $\mathrm{NUV}rK$ diagram. 
  This improvement is probably 
  due to the fact that  
   the high-sSFR subsample is 
  more similar to the late-type galaxies   
  of \citet{Bolzonella2010},  which they identify   using 
  an empirical set of 
   galaxy templates.   We note that also in \citet{Bundy2006} a  difference
  between LD and HD mass function is visible but not
  significant \citep[][Fig.~11]{Bundy2006}. 
   \citet{Mortlock2015},  
   with a combination of photometric redshift samples,  
   conduct a study of environmental effects up to $z\sim2.5$. 
   Their analysis suggests 
   that  massive galaxies at $z<1$  favour denser environment.  
   When they derive the GSMF in this environment 
   they also observe   a flatter low-mass end,  in agreement with our findings.  
   On the contrary, at $z>1$ the 
    GSMFs in low and high densities become very similar.

  In contrast, other studies find no environmental dependency 
  in the stellar mass function of galaxy clusters 
  \citep{Calvi2013,Vulcani2012,Vulcani2013,vanderBurg2013}.  
  The lack of differences in the   
  field vs cluster comparison  
  can be due   
  to the various (local) environments embraced in the 
  broad definition of ``field'' (i.e., a sky region without clusters) 
  that can include single galaxies, pairs, and even galaxy groups.  
  Simulations of \citet{McGee2009} indicate that the majority  
  of cluster members  have been
  accreted through galaxy groups. Other models,  as those 
  used in  \citet{DeLucia2012}, similarly show that a  large fraction of 
  cluster galaxies before  
  belonged to smaller  groups, and were ``pre-processed'' in that environment. 
  Therefore, as much as galaxy groups also 
  contribute  to the stellar mass function in the field,
  the high-mass end is expected to be similar in the  
  two environments. 
  Indeed, when \citet{Calvi2013}
  consider only isolated galaxies, they obtain a stellar 
  mass function that differs from the others. 
  The presence of structures in the field can thus be crucial
  in this kind of analysis.
  
  Also the (global) environment represented by a galaxy cluster 
  includes regions with different local conditions. 
  We note that in 
  \citet{Vulcani2012} 
  the local galaxy density 
  assumes a wide range of values 
  also in clusters. 
  The issue is discussed also 
  by \citet{Annunziatella2014}, who  
  analyse a cluster from the CLASH-VLT survey. 
  They find that the stellar 
  mass function of passive galaxies in the core 
  shows a steeper decrease at low masses,  
  in comparison with passive galaxies in the outskirts of the cluster. 
  In addition, we emphasise that VIPERS is better designed 
  than current cluster surveys to probe $\mathcal{M}>\mathcal{M_\star}$.
  For instance, \citet{vanderBurg2013} have
  12 spectroscopic members in their 10 GCLASS clusters 
  with $11.2<\log(\mathcal{M/M_\odot})<11.6$ 
  and no detection at higher masses; instead, our HD regions contain 
  a few hundreds (spectroscopic) galaxies above 
  $\log(\mathcal{M/M_\odot})=11.2$.
 
  To summarise, the comparison illustrates the advancement  
  VIPERS represents with respect to previous surveys like zCOSMOS 
  or DEEP2: 
  we are now able to 
  robustly discriminate the LD and HD mass functions, 
  finding differences 
  that were not  statistically significant before.  We emphasise 
  that VIPERS has also  more statistical power than current 
  cluster surveys to probe the massive end of the GSMF. 
  Besides that, the fact that our results disagree e.g.~with \citet{Vulcani2012} 
  is related to the different definition of environment. 
  On the other hand, the sample used in this paper 
  spans only  $\sim\!2.3\,\mathrm{Gyr}$ 
  of the history of the universe, whereas  zCOSMOS and DEEP2 have a larger 
  redshift range. Future spectroscopic surveys shall 
  combine high statistics 
  and large cosmic time intervals 
  thanks to the next-generation 
  facilities \citep[especially PFS, the Subaru Prime Focus Spectrograph][]{Takada2014}. They could 
  confirm whether the environmental effects on the GSMF at $z\lesssim1$ 
  (i.e., the enhancement of the high-mass end and the flattening of the 
  power-law slope) 
  vanish at higher redshifts,  as suggested by \citet{Mortlock2015}.

   We can also compare the VIPERS mass functions with 
   those measured in the local universe. In particular, 
   \citet[][hereafter P10]{Peng2010}   
   define the environment of SDSS galaxies 
   as in \citet{Bolzonella2010}, 
   i.e.~in a way similar to ours. They  find:  
  \begin{enumerate}[i.]
  \item  values of $\alpha$ and $\mathcal{M}_\star$  for  
  active GSMFs are the same in the LD and HD regions;
  \item in LD, 
  the stellar mass function of  passive galaxies has the  
  same $\mathcal{M}_\star$ of the active one; 
  \item comparing passive the GSMF in LD and HD regions, the latter 
  have a larger value of $\mathcal{M}_\star$.\footnote{
  In P10, the passive GSMFs are fitted with a double Schechter function  
  Here we refer only to what concerns the primary (most massive) 
  component.  
  }
  \end{enumerate}  
  Thanks to the large volume of the VIPERS sample, 
  and to the high precision of the redshfit measurements, 
  we can verify whether these findings extend to intermediate redshifts.
  We emphasise that at $z>0$ the environmental signatures   
  (i)--(iii)  have not been confirmed yet: several studies provided 
  contrasting clues \citep[cf][]{Bolzonella2010,Vulcani2012,
  Giodini2012,vanderBurg2013,
  Annunziatella2014}.    
    
   With respect to the passive mass functions,  the STY estimator yields   
  larger 
  $\mathcal{M}_\star$ values in the regions of higher density, 
   as stated in (iii). 
  We find such a trend in all three redshift bins 
  (see Table \ref{tab_mstar}).   
  This feature, as we will discuss   
  in Sect.~\ref{Evolution}, can be associated to  
   dry major mergers, which are more likely to happen 
   in the overdense regions. 
  Turning  to the active GSMFs, 
  we observe (i) and (ii) at $z>0.65$. Indeed,   
  the shape of the active GSMF is similar 
  in the two VIPERS environments, 
  since  
  $\alpha$ and $\mathcal{M}_\star$ 
  computed in LD/HD regions are 
  compatible within the errors (note that at $z\sim0.84$ 
  we can compare only $\mathcal{M}_\star$ because 
  $\alpha$ is fixed \textit{a-priori}).  
  Moreover,  $\mathcal{M}_\star^\mathrm{act,LD}$ is 
  consistent with   $\mathcal{M}_\star^\mathrm{pass,LD}$. 
   At $0.51<z\leqslant0.65$, 
   the features (i) and (ii) 
   are not observed any longer. 
  We  argue that  the difficulty in assessing clearly 
   (i) and (ii) is due to the GSMF parametrisation of 
  the active sample, which here is 
  a single Schechter function  (Eq.~\ref{schfun}). 
  Recent work suggests that this is not the optimal choice.   
  For instance, \citet[][GAMA survey]{Baldry2012} observe 
  an excess of blue galaxies 
   at $\mathcal{M}>10^{10}\,\mathcal{M}_\odot$ 
  with respect to their best (single Schechter) fit, 
  with the magnitude of the deviation 
  depending on the colour adopted to classify.  
  We find that, by adopting a double-Schechter model for 
  the active mass function at $z\sim0.6$,  
  the STY fit  produces $\alpha$ and $\mathcal{M}_\star$ 
  that satisfy relations (i) and (ii). However, the uncertainties in 
  this case are larger: given the stellar mass 
  limit of VIPERS,  the slope of the secondary component is not 
  well constrained. 
  In the next Section we will discuss the origin of 
  these GSMF features, already observed in the local universe and 
  now confirmed at $0.5\lesssim z \lesssim1$
  
  \section{Discussion}
  \label{Discussion}

  The  shape of the passive GSMFs is different in the LD 
  and HD environments, and this difference increases going to 
  lower $z$ (see Fig.~\ref{gsmf1}). 
  This can be the result of an environmental-dependent  quenching 
  mechanism, but may also be explained by a different halo mass 
  distribution, or a different assembly history for haloes 
  of similar mass but residing in different regions 
  \citep[see discussion in][]{DeLucia2012}. 
  A similar perspective, looking at the halo environment, 
    has been adopted 
    by \citet{Hearin2015} to explain the so-called ``galactic conformity'' 
    \citep{Weinmann2006}, which is the tendency of satellite galaxies 
    to stay in the same state (star forming of passive)  
    of the central one well beyond the virial radius. 
     Such a sSFR correlation could be linked to the tidal 
    forces that 
    haloes evolving in the same large-scale environment experienced.

  \subsection{Comparison with semi-analytical models}
 \label{Millennium}  

    \begin{figure}
  \includegraphics[width=0.49\textwidth]{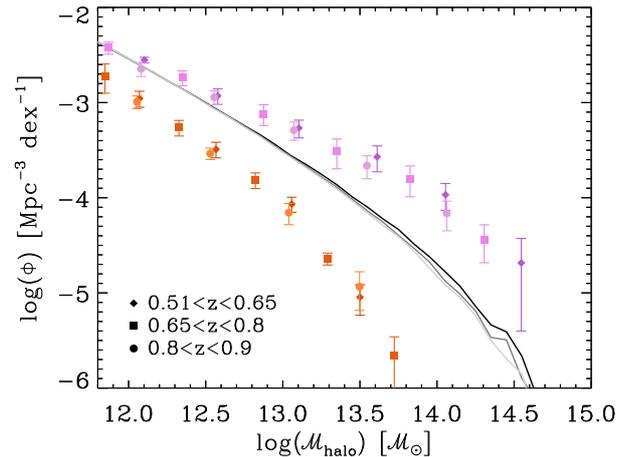}
  \caption{Halo mass function derived from the simulation described in Sect.~\ref{Millennium}, 
  restricted to galaxies  in the HD and LD environment (violet and orange symbols, 
  respectively). Different symbols are estimated in the three redshift bins quoted in the 
  bottom-left corner of the plot, with error bars obtained from the variance among the 10 
  mock catalogues. The mass function of haloes in the entire box 
  ($714\,h_{70}^{-1}\,\mathrm{Mpc}$ side), at snapshots consistent with our 
  redshift bins, is shown as reference 
  with solid lines (darker colour at lower $z$).  
  }
  \label{mill_hmf}
  \end{figure}      
  
  We make use of galaxy simulations to 
  investigate more in detail the two environments we defined. 
  In VIPERS we can exploit 
  a set of 10  light cones, built  
  from the Millennium simulation \citep{Springel2005}. 
  To derive mock catalogues, 
  dark-matter haloes are populated with galaxies 
  by means of the semi-analytic model 
  (SAM)  of \citet[][hereafter DLB07]{DeLucia&Blaizot2007}. 
  For each mock galaxy, rest-frame and apparent magnitudes 
  have been estimated in the same  filter used in the real survey, 
  and the 
  same magnitude cut of VIPERS ($i \leqslant 22.5$) is applied to 
  the simulated catalogue. 
  We add an error to each redshift  
  to emulate observational measurements,  either spectroscopic or photometric 
  depending  whether the object is chosen to be a VIMOS pseudo-target
   by the  slit positioning algorithm.\footnote{The sampling rate is defined 
   as the ratio between the number of spectroscopic pseudo-targets and the 
   whole mock galaxies sample, in bins of redshift.  It is  
   very similar to the TSR of the real survey, 
   while the SSR is 100\%. The statistical weighing factor is 
   therefore $1/\mathrm{TSR(z)}$.}   
  In Appendix \ref{Appendix} and \ref{Voronoi} we use these mock catalogues 
  to test our reconstruction of the  
  density field, 
   together with another  set  realised through the halo 
  occupation distribution (HOD) technique 
  \citep[see][]{delaTorre2013b}. 
  
  The HOD mock galaxies  
  better reproduce VIPERS-PDR1: they cover the same area of 
  the real survey and have  the colour pre-selection applied. 
  The SAM catalogues were prepared at an earlier stage of the survey, so 
  in each of the 10 realisations the effective  area is $4.5\,\mathrm{deg}^2$.   
  The  decline of $N(z)$ at $z\sim0.5$ due to the VIPERS selection function 
  is reproduced by removing objects randomly, 
  irrespective of their $(g-r)$ and $(r-i)$ colours. 
  Nevertheless, the SAM catalogues are better suited 
  to the goal of this Section, containing more physical 
  information.  
  Indeed, the DLB07 model  predicts  
  galaxy properties such as stellar mass, SFR, colours 
  at different redshifts, 
  in addition to the apparent magnitudes mentioned above; 
  on the contrary, galaxy stellar mass and SFR 
  are not available in the HOD catalogues.

  In these Millennium light-cones  we identify HD and LD regions  
  by means of the same method used with real data 
  (see Sect.~\ref{Environment definition} and Appendix \ref{Voronoi}).  
  In principle, this means that environmental effects predicted by DLB07 can   
  be straightforwardly compared to those found in VIPERS.   
  However  the LD/HD environments in the simulation may correspond 
  only roughly to the regions delimited 
  in the real survey, for several reasons. 
  First, the volume-limited ($M_B<20.4-z$) 
  tracers used to estimate $\delta$  in the simulation 
   may have different number density 
  and clustering. As highlighted in 
  \cite{Cucciati2012b}, at intermediate redshifts 
  the $B$-band luminosity function     
  shows an excess of bright late-type galaxies 
  in the DLB07 model with respect to VVDS data, 
  while early-type  galaxies at $M_B<M_B^\star$ are underpredicted.   
  Moreover, we are aware that for  the most luminous and massive galaxies 
  the two-point correlation function of VIPERS is slightly higher  than DLB07  
  on scales $\gtrsim 7\,h_\mathrm{70}^{-1}\,\mathrm{Mpc}$  \citep{Marulli2013}.
  This is expected, as the  $\sigma_8$  parameter, set 
  by the first-year analysis of the Wilkinson Microwave Anisotropy Probe 
  \citep[WMAP1,][]{Spergel2003} and adopted in the Millennium simulation, 
  is larger than more recent measurements from WMAP9 and Planck-2015  
  \citep[][]{Hinshaw2013,Planck2015-13}. 
   We discuss these differences also 
   in Appendix \ref{Voronoi}.  
   Further  investigations have been carried out in Cucciati et al.~(in prep.). 
  Overall,  those tests show that 
  structures (and voids) in the Millennium simulations grow earlier 
  than those in the observed universe, 
  and the volume occupied by the HD (LD) regions is smaller (larger).

  Nevertheless, the under- and over-densities in our light cones 
  still represent two opposite environments that we can contrast,     
  e.g.~by looking at  their underlying  
  dark matter content.   
    Figure~\ref{mill_hmf}  shows 
    the mass distribution of  haloes hosting either LD or HD galaxies. 
    In all the redshift bins, the number density of HD haloes is higher 
    than the LD ones. The distribution of the former has a flatter slope, 
    with a higher fraction of massive haloes:  
    those with $\mathcal{M}_\mathrm{halo} \gtrsim 10^{13.5}\,\mathcal{M}_\odot$ 
    are not found in the opposite, low-dense environment.  This excess  
    is a clear indication that our environment 
    reconstruction classifies as HD regions rich galaxy groups and galaxy 
    clusters. 
    These results are in agreement with \citet{Fossati2015}, who find 
    similar correlations between  local galaxy density and halo mass in 
    a thorough study of galaxy environment. We also highlight that the 
    halo number density starts to be higher in HD at masses of 
     $10^{12}$--$10^{12.5}\,\mathcal{M}_\odot$. Haloes in this bin should includes 
     almost 50\% of galaxies with $\mathcal{M}>10^{11}\,\mathcal{M}_\odot$, as 
     found by \citet{Popesso2015}.\footnote{We note that both  
    \citet{Fossati2015} and \citet{Popesso2015} use SAMs from the same 
  ``family'' of DLB07, implemented on a new run of the Millennium simulation.}

  The difference observed between LD and HD 
  in the high-mass end of the GSMF (Fig.~\ref{gsmf1}) can be  
  interpreted, at least partly, as a reflection of the mass
  segregation of dark matter. 
  In hierarchical models, massive haloes 
  preferentially populate the densest regions 
  \citep[e.g.][]{Mo&White1996}, and the correlation between
  halo mass and galaxy stellar mass 
  produces in turn a concentration of
  massive galaxies in the HD environment 
  \citep[cf][]{Abbas2005,Abbas2006,Scodeggio2009,delaTorre2010}. 
    This gives an idea about how intrinsic properties of the galactic systems 
  are entangled with the classification  of  their local  environment via halo mass, 
  without any 
  solution for the ``nature'' vs ``nurture'' dilemma.   
   This picture is  consistent with the mass segregation 
   observed by \citet{vanderBurg2013} in the GCLASS clusters at 
   $z\simeq1$. They normalise their stellar mass function  
   by estimating the total mass (baryons and dark matter) 
   contained within the virial radius of each cluster. On the other hand, 
   their GSMF in the  UltraVISTA field is normalised by multiplying its 
   volume by the average matter density of the Universe. 
   After such rescaling, the authors find that the stellar 
   mass function is higher in the clusters than in the field 
   \citep[see][Fig.~8]{vanderBurg2013}.

  We can also derive the stellar mass function of SAM galaxies 
  in LD and HD environments. 
  We already know (see D13) that the DLB07 model 
  overestimates the GSMF 
  low-mass end of the VIPERS field, and 
  shows minor tension at higher masses.  
  The same weaknesses are present in more recent SAMs 
  \citep[see e.g.][]{Fontanot2009,Cirasuolo2010,Guo2011,Maraston2013,Lu2014} 
  and also  in hydrodynamical simulations.  
  Furthermore, discrepancies  arise  
   because of the error sources in the observations 
  \citep[e.g.~systematics in stellar mass estimates, see][]{Marchesini2009,Bernardi2013}. 
  Most importantly, the LD and HD regions traced in the simulation, although having  
  the same meaning of the real ones, are different e.g.~in terms of occupied volume 
  (see discussion above). For this reason we renormalise each GSMF to unity 
  number density (as previously done in Fig.~\ref{gsmf1}). 
  
\begin{figure}  
 \includegraphics[width=0.99\columnwidth]{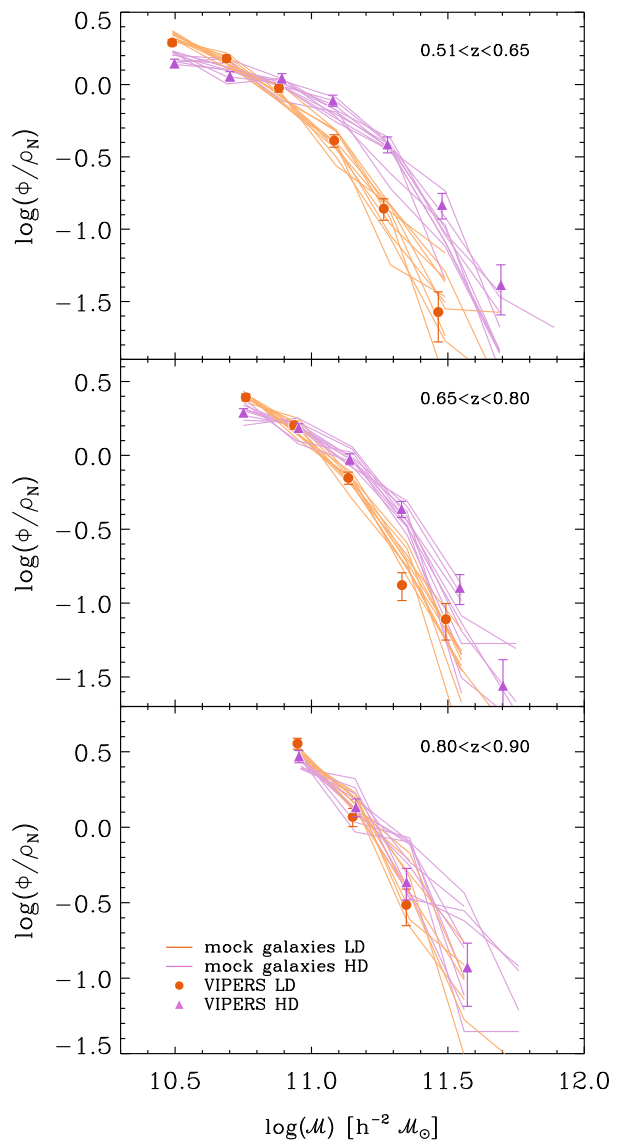}
 \caption{Stellar mass functions  of mock galaxies built from the Millennium simulation 
  through the semi-analytical model of \citet{DeLucia&Blaizot2007}.  The 10 mock 
  realisations correspond to the solid lines (orange and violet for LD and 
  HD regions respectively) while symbols with error bars show the 
  GSMF of VIPERS in the two environments (the same as Fig.~\ref{gsmf1}). 
  All the mass functions are plotted starting from the completeness limit 
  ($\mathcal{M}_\mathrm{lim}$) at that redshift. They are  
  obtained by means of the $1/V_\mathrm{max}$ method, 
  rescaled  to have the same number density $\rho_N$  when integrating 
  $\Phi(\mathcal{M}$ at $\mathcal{M}>\mathcal{M}_\mathrm{lim}$. 
      }
      \label{mill_smf}
\end{figure}  
  
  The shape of the different GSMFs are compared in  Fig.~\ref{mill_smf}. 
  In both environments, at each redshift bin, 
  the shape of the mock GSMF is similar to the observed one   
  after convolving SAM stellar masses  with a Gaussian of dispersion 0.2\,dex, 
  to reproduce observational uncertainties. The 0.2\,dex width 
  has been chosen as an arbitrary value representing the typical scatter in 
  the SED fitting estimates \citep[see e.g.][]{Mobasher2015}. We note that a different value 
  \citep[e.g.~0.25\,dex, as in][]{Guo2011} may result in a worse agreement 
  with data.  Aware of this potential bias, we note that it would not 
  remove the difference emerging between HD and LD regions in the 
  simulation. Indeed, the main finding in this Section is  that mock GSMFs 
  show the same increase  
  of the high-mass end in the densest environment, as found in VIPERS.

  In addition to this, the model hints how the low-mass slope changes 
  as a function of environment, at least for the GSMF at $0.51<z\leqslant0.65$ where the 
  mass range probed is the largest.  
    Looking at the central galaxies (as defined according to the merger tree)  we note that 
    about half of those living the HD regions are central of a sub-halo already inside a larger structure, 
    while in the LD regions most of them are  ``isolated'' central. 
   Also the number of satellite galaxies, i.e.~those 
   embedded in  another galaxy halo,  
   increases as a function of $\delta$: 
    the HD satellite fraction  is  a factor $\sim2$ higher than the one in LD, reaching about 20\% at 
    $\log(\mathcal{M/M}_\odot)\sim 10.6$ and going down to zero at  
    $\log(\mathcal{M/M}_\odot) > 11$.  
    Also the  fraction of recent mergers (i.e., mergers between two consecutive 
    timesteps) is $\sim2$ time larger in the HD regions.   
    This can explain the flatter profile of the GSMF with respect to the LD regions.  
    The relevance of mergers is discussed, with a different approach, also in the next Section.

 \subsection{An empirical approach}
  \label{Evolution}
 
  We use VIPERS data to test  the empirical description
  of galaxy evolution proposed by P10, in which   
  the galaxy number density changes as a function of 
  $\mathcal{M}$, SFR, and  environment. 
  Three observational facts are fundamental in  P10: 
  \begin{enumerate}[1)]
  \item the stellar mass function of star-forming galaxies 
  has the same shape at different redshifts 
  \citep[i.e., $\alpha$ and $\mathcal{M_\star}$  
  are nearly constant, see e.g.][]{Ilbert2010}, with
  little increase in normalisation moving towards lower 
  redshifts;  
  \item there is a tight relation between SFR and stellar mass for 
  star-forming galaxies  (the so-called ``main sequence'')
  with $\mathrm{SFR}\propto\mathcal{M}^{1+\beta}$ 
  \citep[e.g.][]{Noeske2007,Elbaz2007,Daddi2007};
  \item average sSFR can be parametrised with respect to
  stellar mass and redshift/cosmic time 
  \citep[][and references therein]{Speagle2014}, 
  while it is independent of environment 
  \citep[P10;][]{Muzzin2012,Wetzel2012}.   
  \end{enumerate}
  In spite of the large 
  consensus in the literature, we caution that
  these three findings have been  established only recently:     
  new data may be at odds with them, 
  bringing into question the basis of \citeauthor{Peng2010} work. 
  For instance, \citet{Ilbert2015} show that 
  $\log(\mathrm{SFR})\propto -\mathcal{M}\log(\mathcal{M})$ is a better 
  parametrisation than 2), at least for their $24\mu m$ selcted sample.  

  The keystone of P10 description is that   
  two mechanisms can regulate the decline 
   of star formation; they are named  
  \textit{mass quenching} and \textit{environment quenching} 
  as they depend respectively on $\mathcal{M}$ and $\delta$. 
  In first approximation, the evolution of the GSMF can be 
  parametrised by the two mechanisms only.  
  As we shall show below, other processes 
  are needed in the HD regions. 
  Using data from local Universe
  \citep[SDSS-DR7,][]{Abazajian2009} and at $z\sim1$
  \citep[zCOSMOS,][]{Lilly2007} the authors argue that
  mass and environment quenching are fully separable. 
  The effect of both can be expressed 
  analytically; in particular the mass quenching rate is 
   \begin{equation}
     \lambda_m  = \frac{\mathrm{SFR}}{\mathcal{M_\star}} 
     = \mu \mathrm{SFR} \:,
    \label{massquen}
   \end{equation}
  where $\mathcal{M_\star}$, namely the Schechter parameter of the
  star-forming mass function, is  constant ($\mathcal{M_\star} \equiv 
  \mu^{-1} \simeq 10^{10.6}\,\mathcal{M_\odot}$, according to 
  observations). 
  Equation~(\ref{massquen}) can be regarded as 
  the probability of a galaxy to become passive via mass quenching. 
  This is the simplest analytical form that satisfies 1)-3) but alternative, more complex 
  formulations cannot be excluded.  
  
  The empirical laws of P10 
  do not shed light on the physical processes
  responsible for quenching but  
  describe its  characteristics.   
  In \citet{Peng2012} mass and environment quenching are linked 
  to halo occupation. In this view, central galaxies are subjected to the former, 
  which is analogous to the ``internal quenching'' described in 
  other papers \citep[e.g.][and reference therein]{Gabor2010,Woo2012},  
  while environment quenching is the preferred channel of satellite galaxies. 
  This distinction however is not clear-cut because satellite galaxies can 
  spend a significant fraction of their life as centrals, before being accreted 
  into another halo  \citep[see e.g.][]{DeLucia2012}. 
  Moreover \citet{Knobel2015}, using the same SDSS group catalogue 
  of  \citet{Peng2012}, show that the  central vs satellite  
  dichotomy disappears when excluding isolated galaxies from the 
  sample of central galaxies (i.e., central galaxies in groups are 
  affected by the environment in the same way as satellites).

  With these simple prescriptions,  
  it is possible to  
  reproduce several statistics of galaxies across cosmic time.
  In P10, the authors 
  generate a galaxy sample   
  at $z=10$, with a primordial stellar mass function  
  that follows a power law, 
  and they evolve it
  down to $z=0$.   
  That mock sample has very simple 
  features, e.g.~active galaxies 
  form stars at a constant level 
  that is given by the 	
  sSFR$(z,\mathcal{M})$ parametrisation 
  of \citet{Pannella2009}.   
  At any epoch, a fraction of blue galaxies become 
  red, proportionally to mass and environment quenching 
  rates. 
  This picture does not include the birth of 
  new galaxies.   
  
  Here, we do not make use of mock galaxies, rather we 
  start from  the 
  observed stellar mass function  in 
  a given $z$-bin 
  and  ``evolve'' it to a lower redshift following the prescriptions of P10. 
  Then, we compare such an ``empirical prediction'' of the  
  GSMF with our data.   
   
\begin{figure}
  \includegraphics[width=0.97\columnwidth]{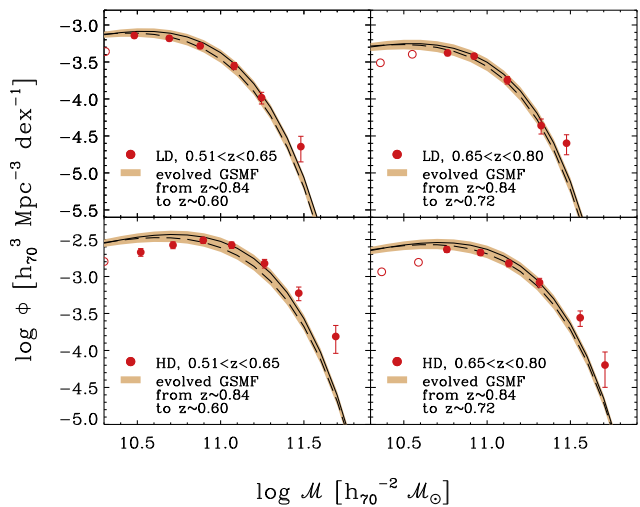}
  \caption{Comparison between the GSMFs constructed with the  
  P10 recipe and the VIPERS data. 
  In each panel, red filled circles are the $1/V_\mathrm{max}$ points 
  (with Poissonian errors) of the VIPERS 
  passive mass function, in the redshift bin and environment indicated 
  in the legend; 
  lines and shaded area represent the evolution 
   of the GSMF observed at $0.8<z<0.9$, 
  down to the same redshift of the plotted data points.  
  Applying the   
  quenching description of P10, we obtain two different estimates 
  if we use  the original  sSFR$(z,\mathcal{M})$ parametrisation 
  of P10 (solid line), or the function provided in 
  \citet[][dashed line]{Speagle2014};  
  a further error is introduced to account for the uncertainties in the 
  integration (see Eq.~\ref{massquen}), giving the final width of the 
  shaded area.
 }
  \label{mfev}
\end{figure}

   In the LD regions, the fraction 
  of VIPERS active galaxies that migrate into the passive mass function  
  is assumed  $\propto \lambda_\mathrm{m}$, i.e.~it is 
  determined by mass quenching only. 
    To evaluate the fraction of new quenched galaxies, one has to insert a 
   functional form of the
   specific SFR, generally speaking sSFR$(z,\mathcal{M})$, into 
   Eq.~\ref{massquen}. The function chosen by P10 (their Eq.~1) 
   comes from \citet{Pannella2009}. 
   From such a definition of quenching rate, 
   it follows that, in a given mass bin centred in $\mathcal{M}_\mathrm{b}$, 
   the galaxy number density evolution is 
   \begin{align}
   \Phi_\mathrm{pass}(z_2) \,=&\, \Phi_\mathrm{pass}(z_1) + 
   \int_{t(z_1)}^{t(z_2)} \Phi_\mathrm{act}(z) 
   \lambda_\mathrm{m} \,\mathrm{d}t  \nonumber \\
    =&\, \Phi_\mathrm{pass}(z_1) + \tilde{\Phi}_\mathrm{act} \,
    \mu  \int_{z_1}^{z_2} \mathcal{M}_\mathrm{b} \,
    \mathrm{sSFR}(z,\mathcal{M}_\mathrm{b}) 
      \,\mathrm{d}z \;. 
      \label{phievo}
   \end{align}
  In the Equation, the  GSMF of the active sample 
  is  constant ($\tilde{\Phi}_\mathrm{act}$) 
  between $z_1$ and  $z_2<z_1$, regardless of the 
  environment in which it is computed.  
  This assumption is supported both by our data (see Fig.~\ref{mfcfvmax}) 
  and other studies \citep[e.g.][]{Pozzetti2010,Ilbert2013};  
  $\tilde{\Phi}_\mathrm{act}$ is determined by averaging the 
  $\Phi_\mathrm{act}$ estimates at $z_1$ and $z_2$. 
  We apply Eq.~(\ref{phievo}) in the LD environment, 
   evolving data at $0.8<z\leqslant0.9$  down to 
   $\langle z\rangle=0.72$ and 
   $\langle z\rangle=0.6$.  
   The resulting passive GSMFs, built under the action of 
   mass quenching only, are consistent with those 
   observed in the corresponding redshift bins 
   (see Fig.~\ref{mfev}, upper panels).  
   We repeat the procedure 
   starting from $0.65<z<0.8$, finding a good agreement 
   at $\langle z\rangle=0.6$  (this comparison is 
   not shown in the Figure). 
   
  The major uncertainty in this technique  is related to 
   SFR-$\mathcal{M}$ relation. To quantify the 
   impact of different  parametrisations, we also used, 
   instead of the equation provided in P10, the ``concordance function''   
   obtained by \citet{Speagle2014} fitting   
   data of 25 studies from the literature (see their Eq.~28). 
   We also estimate the uncertainty related to  
   $\tilde{\Phi}_\mathrm{act}$ by replacing it 
   with upper and lower values of 
   $\Phi_\mathrm{act}(z_1)$ and $\Phi_\mathrm{act}(z_2)$ 
   respectively. We note that keeping 
   the active mass function  fixed 
   introduces a much smaller uncertainty 
   with respect to 
   the sSFR$(z,\mathcal{M})$ parametrisation. 
   Another approximation in the procedure is 
   that galaxies do not change 
   environment as time goes by. 
   This assumption is appropriate  
   in the time interval we probe, as we verified following the 
   evolution of mock galaxies in the simulations of Sect.~\ref{Millennium}.

\begin{figure}
  \includegraphics[width=\columnwidth]{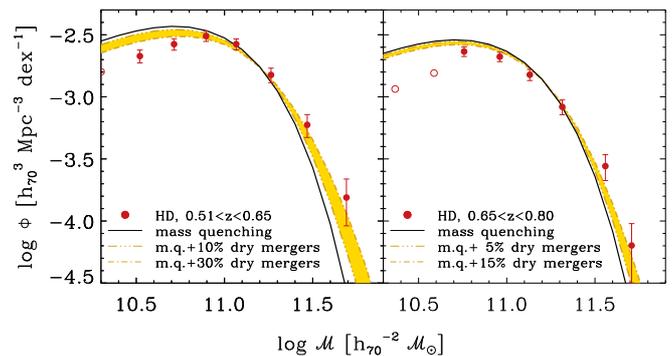}
   \caption{Evolution of the passive mass function in the HD environment, 
    including 
    dry mergers. The solid line in each panel is the 
    predicted GSMF in the HD environment, 
    as in Fig.~\ref{mfev}, 
    assuming mass quenching only and the sSFR parametrisation of P10; 
    yellow shaded area is the GSMF modified by dry mergers, whose 
    percentage ranges from 5--10\% (triple-dot-dashed line) 
    to 15--30\% (dot-dashed line) depending 
    on the redshift bin.  In each $z$-bin, 
    red circles are the $1/V_\mathrm{max}$ 
    estimates (with Poissonian errors) 
    of the stellar mass function of the VIPERS passive galaxies 
    (symbols are filled above the 
    completeness limit $\mathcal{M}_\mathrm{lim}^\mathrm{pass}$). }
    \label{mfevdry}
\end{figure}

  We apply Eq.~(\ref{phievo}) also in the HD regions.
  We emphasise that in this case there should be a combined effect of 
  both mass and environment quenching mechanisms. However, P10 show that 
  the former is more effective 
  at $\log(\mathcal{M/M_\odot})<10.5$, and 
   therefore negligible  
  in the VIPERS stellar mass range. The main difference 
  with respect to LD, instead, is that  
  after  becoming passive, galaxies in the overdensities 
  have higher chance to merge. 
  We will show that such  dry mergers are crucial to modify the shape of 
  the passive GSMF. 
   In fact, a description which accounts for 
   mass quenching only 
   does not reproduce well 
   the passive mass function of HD galaxies  
   (Fig.~\ref{mfev}, lower panels). 
      Dry mergers produce a  redistribution of the stellar mass 
   in the simulated GSMF, which is now more consistent 
   with the observed one (Fig.~\ref{mfevdry}).  
   We add this `post-quenching' ingredient (i.e.~dry merging) 
   through the scheme described below. 
   
   P10 assume a simple model in which 
   part  of the passive population  
   merges with 1:1 mass ratio. 
   Similar prescriptions are used also in the ``backward 
   evolutionary model'' of 
   \citet{Boissier2010}.  
   Both  P10 and \citet{Boissier2010} find that 
   dry major mergers  enhance the exponential tail 
   of the passive GSMF, and make  $\mathcal{M}_\star$ 
   increase with respect to the LD environment. 
      They also consider minor 
   mergers fully negligible in the GSMF evolution, at least at   
   $\mathcal{M}\geqslant 10^{10}\,\mathcal{M}_\odot$,   
   \citep[see also][]{Lopez-Sanjuan2011,Ferreras2014}.  
   In our analysis, we introduce dry (major) mergers in the evolution of 
   $\Phi_\mathrm{pass,HD}$, assuming that two 
   objects  
   in the same bin of $\log\mathcal{M}$  can merge together 
   without  triggering relevant episodes of star formation 
    \citep[e.g.][]{DiMatteo2005,Karman2015}.  
   We set the fraction of galaxies 
   undergoing a 1:1 merger 
   to be equal to $f_\mathrm{dry}(z)$, 
   with no dependence on the stellar mass of the 
   initial pair \citep[cf][]{Xu2012a}. 
   An estimate of  $f_\mathrm{dry}(z)$ is 
   inferred by \citet{Man2014} by counting galaxy pairs 
    with stellar mass ratio less than  $1:4$.\footnote{\citeauthor{Man2014}  
    show that their merger rate is  
   suitable to study dry mergers, e.g.~it is 
   consistent with that of gas-poor galaxies 
   in the simulations of  \citet[][]{Hopkins2010a}. 
   Moreover 
   the authors, performing their analysis   
   on the COSMOS field, 
   can be compared to several previous studies \citep[e.g.][]{deRavel2011,
   Xu2012a,Lopez-Sanjuan2012}, with which they are in  
   fairly good agreement.  
   P10 
   use the merger rate derived by \citet{deRavel2011} 
   for the zCOSMOS galaxies.}

  The merger rate of \citet{Man2014} leads to a merger 
  fraction  $f_\mathrm{dry}=5_{-2}^{+3}\%$ 
  from $\langle z\rangle=0.84$ to 0.72, 
  and $f_\mathrm{dry}=10_{-4}^{+6}\%$ 
  from $\langle z\rangle =0.84$ to 0.6. 
  Since they are averaged over the general COSMOS field, 
  these values can get  $\sim$2--3 times   
  higher in  HD  environments 
  \citep[][see also \citealp{Lin2010,Lotz2013}]{Kampczyk2013}.  
   For this reason we test a range of $f_\mathrm{dry}$ values: 
    from 5 to 15\% in the  time span from $\langle z\rangle=0.84$ to 
    0.72 ($\sim$0.7\,Gyr) 
    and 10--30\% 
     from  redshift $0.84$ to 0.6  (i.e., across  $\sim$1.4\,Gyr).  
   As stressed above, dry merging is the key element to reconcile the simulated GSMF 
   with the observed one (Fig.~\ref{mfevdry}).  
   Nevertheless, a 1$\sigma$ difference remains at 
   $\log(\mathcal{M/M_\odot})\simeq10.4$. 
   Together with the ($<1\sigma$) difference  at the high-mass 
   end, this overestimate may suggest that the impact of mergers in the 
   densest regions of VIPERS could be even larger than what we assumed.  
   At the same time,  we cannot exclude that the explanation for these (minor) tensions  
   resides in the simplicity of our parametrisation. Indeed the result depends on the 
   model used to describe the evolution of these massive galaxies. 
   For example, central ones could grow significantly by
  (multiple) accretion of satellites. Since our sample does not
  distinguish between satellite and central galaxies, we could not
  test this scenario. 

  \section{Conclusions}
  \label{Conclusions}
 
  The large volume probed, along with the accuracy 
  of redshift measurements, make VIPERS the ideal survey 
  to study environmental effects at intermediate redshifts. 
  We reconstruct the local density field (\citealp[][]{Cucciati2014a}; Cucciati et al., in prep.) 
  and identify galaxies embedded in under- and over-densities.  
  We estimate the volumes occupied by 
  such LD and HD regions, finding that they represent 
  nearly $50$ and $7\%$ of the total  comoving volume of the survey.   
  Thanks to the volume reconstruction, we can properly compute  
  the number density of galaxies in these two opposite environments, 
  and compare the GSMFs   
  at different epochs.    
  
  The stellar mass function of LD galaxies  is nearly constant 
  in the redshift range $0.51<z<0.9$, while a significant 
  evolution is observed in the HD regions. 
  Moreover, we find that the VIPERS stellar mass function     
   has a shape that depends on the environment, with 
   a higher fraction of massive objects in the over-densities. 
   Interestingly, our approach is complementary to the other VIPERS studies  
  that show 
  the increase of the galaxy bias as a function of $\mathcal{M}$  
  \citep[e.g.][]{Marulli2013}.   
   Despite our completeness limit 
   ($\mathcal{M}_\mathrm{lim}\gtrsim 10.4$  
   at $z\sim 0.6$) we also find hints that the 
   low-mass end of the  GSMF is flatter in the HD regions, 
   with a particular decrement of the passive sample.  
   This marginal effect could be robustly assessed once the final VIPERS 
   catalogues ($\sim90 000$ spectra) will be available. 
  
   The LD vs HD variance is quantitatively 
   described by the \citet{Schechter1976} parameters: 
   the    $\alpha$-$\mathcal{M}_\star$ likelihood contours  from the STY fit  
   show a significant difference between the two environments. 
  In particular, the enhancement of the GSMF massive end 
  is well described by  $\mathcal{M}_\star$, which increases 
  by $\sim0.25$\,dex in   the HD regions 
  (namely $0.24\pm0.12$, $0.27\pm0.15$, and $0.21\pm0.11$\,dex at $z\sim0.6$,
   $0.72$, and $0.84$ respectively).    
  Such a difference  
  remains visible  when considering the 
  active or passive sample only. 
  An environmental imprint in the stellar mass function has already 
  been observed in the local Universe 
  \citep{Baldry2006,Peng2010}. With VIPERS, it becomes evident 
  for the first time also at $z\gtrsim0.5$.

  We investigate these environmental trends  
  by using 10 mock catalogues 
  derived from the Millennium simulation.  Galaxies are 
  simulated  following the prescriptions of 
  \citet{DeLucia&Blaizot2007} and the survey design is reproduced 
  to make these catalogues similar to VIPERS. 
  In this way we were able to define galaxy environments 
  as done in the real survey.  
  The different slope of the low-mass end is observed 
  also in the mock GSMF,  
  and can be associated  to a larger  
  number of merger events where the local galaxy density is higher. 
  Looking at the exponential tail of the mock GSMF, 
  the higher number density of $\mathcal{M}>\mathcal{M}_\star$ 
  galaxies in the HD regions is linked  to a large amount of 
  haloes with $\mathcal{M}_\mathcal{h}>10^{13}\,\mathcal{M}_\odot$.     
  Such massive haloes are absent in the LD sample. 
  As a result, both satellite and merger fractions increase when 
  selecting denser environments.  
  To summarise, our classification based on the galaxy density contrast 
  corresponds to a discrimination  in halo properties, highlighting  the 
   ambiguity of  the ``mass vs environment'' dichotomy \citep[see][]{DeLucia2012}.

  We find that the difference between LD and HD 
  mass functions decreases  
  from  $\langle z \rangle =0.60$  to  $0.84$. 
  The trend is expected to continue at higher redshifts,  
  where the massive haloes that characterise our densest  
  environment have not collapsed yet. 
  We can connect our results to  
  the analysis of \citet{Mortlock2015}, in which the GSMF at 
  $1<z<1.5$ does not change when computed either in 
  high and low densities (even though the large uncertainties 
  could hide some minor environmental effect).   
  This change can be linked to the different conditions of cosmological structures 
  in the earlier stages of the universe, with group environment being 
  more effective at $z<1$. 
  
  We also experimented the  
  empirical description of \citet{Peng2010}, 
  in which the stellar mass function of passive galaxies evolves 
  under the combined effect of mass and environment quenching.      
  Differently from other studies, we use this approach in a self-consistent way:   
  we evolve the observed mass function at each redshift bin considered 
  in our study, and compare the expectation to the GSMF observed at the lower redshift bin.
  Our results show that the measured evolution of the GSMF in low density regions 
  is consistent with a model in which galaxy evolution is dominated by internal physical 
  processes only (``mass quenching'' in the formalism by \citeauthor{Peng2010}). 
  For high density regions, however, additional processes have to be considered 
  to explain the evolution of the massive end of the GSMF. 
  In particular, we demonstrate that the observed evolution can be explained 
  by including the effect of dry mergers.    
  
  We stress that our survey has the capability to shed light 
  on the role of mergers in shaping the GSMF, e.g.~tackling the 
  problem of sample variance highlighted by \citet{Keenan2014}. 
  Moreover, in the redshift range 
  of our survey, merging events are more frequent than in the local universe 
  \citep[][but see also outcomes from state-of-the-art simulations in 
  \citealp{Rodriguez-Gomez2015}]{Lopez-Sanjuan2012}. 
  In our study,  
  both semi-analytic modelling and empirical approach highlighted the 
  importance of mergers in the large-scale dense environment. 
  Future analyses  relying on  the 
  final $24\,\mathrm{deg}^2$ release of VIPERS shall complement the 
  present results, providing further details about galaxy-galaxy interactions.

\begin{acknowledgements}
We acknowledge the crucial contribution of the ESO staff for the management of service observations. In particular, we are deeply grateful to M. Hilker for his constant help and support of this program. Italian participation to VIPERS has been funded by INAF through PRIN 2008 and 2010 programs. 
OC acknowledges the support from grants ASI-INAF I/023/12/0 ``Attivit\`a relative alla fase B2/C per la missione Euclid''. 
 LG and BRG acknowledge support of the European Research Council through the Darklight ERC Advanced Research Grant (\# 291521). 
 OLF acknowledges support of the European Research Council through the EARLY ERC Advanced Research Grant (\# 268107). AP, KM, and JK have been supported by the National Science Centre (grants UMO-2012/07/B/ST9/04425 and UMO-2013/09/D/ST9/04030), the Polish-Swiss Astro Project (co-financed by a grant from Switzerland, through the Swiss Contribution to the enlarged European Union). RT acknowledges financial support from the European Research Council under the European Community's Seventh Framework Programme (FP7/2007-2013)/ERC grant agreement n. 202686.  
EB, FM and LM acknowledge the support from grants ASI-INAF I/023/12/0 and PRIN MIUR 2010-2011. LM also acknowledges financial support from PRIN INAF 2012. YM acknowledges support from CNRS/INSU (Institut National des Sciences de l’Univers) and the Programme National Galaxies et Cosmologie (PNCG). Research conducted within the scope of the HECOLS International Associated Laboratory, supported in part by the Polish NCN grant DEC-2013/08/M/ST9/00664.
\end{acknowledgements}

\begin{appendix}

\section{Tests on the $1+\delta$ distribution}
\label{Appendix}

  In Sect.~\ref{Environment definition} we associated 
  VIPERS galaxies to  LD or HD environments by means of 
  their density contrast $\delta$. Specifically, 
  galaxies with $\delta<0.7$ are 
  assumed to be in LD region, while 
  HD galaxies are those with $\delta>4$.  
  For sake of clarity  
  we dub these thresholds  $\delta_\mathrm{LD}$ and 
  $\delta_\mathrm{HD}$. 
  Their respective values correspond to the 25th and 
  75th percentiles  of the 
  $\delta$ distribution, which can be computed 
  at various redshifts ($0.51<z\leqslant0.65$, $0.65<z\leqslant0.8$, 
  $0.8<z\leqslant0.9$) and 
  in W1 and W4 separately.
  The final thresholds we adopted 
    ($\delta_\mathrm{LD}=0.7$,  
  $\delta_\mathrm{HD}=4$) 
  are obtained by averaging the percentiles 
  obtained in each bin. 
  In this Appendix,  we justify the choice of 
  using constant values  
 despite the small  
  variations among 
  different redshifts and  fields 
  (see Fig.~\ref{d1d4lim}).

  First of all, we verify that the 
  absence of selection effects in the computation. 
  Even though the selection of our spectroscopic targets, 
  described through TSR, SSR, and CSR 
  (Sect.~\ref{Dataspec}), does vary with redshift, this is not 
  the case for the mass-selected sample 
  ($\log(\mathcal{M/M_\odot})>10.86$) we use as a proxy of the 
  density field. The statistical weights of these galaxies are 
  nearly constant from $z=0.51$ to 0.9.

  Some variation of $\delta_\mathrm{LD}$ 
  and $\delta_\mathrm{HD}$ should be 
  due to statistical fluctuations,
  since we are sampling a  nearly-Gaussian  distribution  
  \citep[][]{DiPorto2014}   
  with a limited number of  objects. 
  In fact, each $z$-bin 
  contains only galaxies  that were spectroscopically observed,   
  and the  $\delta$   ranking is sensitive to this incompleteness. 
  From this perspective, the survey selection originates 
  some amount of scatter: datasets 
  drawn from the same 
  galaxy parent population    
  can yield different quartile values 
  just because they populate in different 
  ways the tails of the original density distribution. 
 
  To verify this hypothesis, 
  we perform a Monte Carlo simulation. 
  First, we divide the VIPERS sample  in the 
  three $z$-bins mentioned above, 
  keeping W1 and W4 separate. 
  In each bin, and for both fields individually, 
  we derive a PDF from the observed $\delta$ distribution. 
  We extract 100\,000 
  times the same number of objects as it 
  has been observed in VIPERS, 
  and assign to these  fake galaxies  
  a density contrast according to the reconstructed PDF. 
  In other words, this task consists in reproducing many times 
  the plot shown in Fig.~\ref{d1d4lim}, as it would appear 
  if we targeted different galaxies from the parent photometric 
  sample (every time with the same sampling rate). 
  The quartiles resulting from each realisation have 
  a scatter of the
  order of 10--15\% around the mean value.

  Another reason for the fluctuations of $\delta_\mathrm{LD}$ 
  and $\delta_\mathrm{HD}$ 
  could be cosmic variance. In this case, it is not the subsample 
  of observed objects to vary but the density field itself, 
  e.g.~because of field-to-field variations in large-scale
  clustering \citep[][and references therein]{Moster2011}.  
  In VIPERS, thanks to its large volume, 
  this effect is generally small, as shown in D13 and \citet{Fritz2014}.
  To estimate the impact of cosmic variance on our definition 
  of environment, we use two sets of simulations, each one 
  consisting in 10 independent mock galaxy catalogues. 
  The first set originates from  the halo occupation distribution (HOD)
  modelled by \citet[][see also the description in \citealp{Cucciati2014a}]{
  delaTorre2013b}.\footnote{The other mock catalogues,
   built according the semi-analytical model (see Sect.~\ref{Millennium}),  
   are not used here because they cover   
   a single sky region of $7\times1\,\mathrm{deg}^2$.} 
   
   To do that, we started from mock catalogues that have 
   100\% sampling rate, no masked area, and 
   galaxy redshifts without observational errors (i.e., they are 
   cosmological redshifts perturbed by peculiar velocities).
   We referred to them as ``reference'' mock catalogues. 
   We manipulated them    
   to reproduce the VIMOS footprint, and added redshift 
   measuring errors to have the correct percentages of 
   $z_\mathrm{phot}$ and $z_\mathrm{spec}$  
   (``VIPERS-like'' mock catalogues). 
   We estimate galaxy density contrast 
   (through the projected 5NN, as described in Sect.~\ref{Density contrast}) 
   and consequently its distribution, in the three $z$-bins used in 
   this work. 
   Among the 10 ``VIPERS-like'' realisations using HOD, 
   the 25th (75th) percentile  
    that determines the LD  (HD) environment has
   $\sim5\%$ ($\sim10\%$) scatter. 
   This outcome implies that the
   LD and HD thresholds in real data vary
   also because of cosmic variance. In the HOD mock catalogues 
   the galaxy luminosity in the B band is available. Assuming an average 
   $\mathcal{M}/L_\mathrm{B}$ ratio, we estimated the fractional error 
   due to cosmic variance in each bin of stellar mass of 
   the total GSMF shown in Fig.~\ref{gsmf1}. 

   In conclusion, the percentiles we estimated for VIPERS,  
   in its two fields and 
   within three different $z$-bins,   
   spread over a range comparable 
   to  the one  resulting in simulations. 
   Undersampling of the $\delta$ distribution 
   and cosmic variance are the major 
   responsible for these fluctuations, which 
   are small enough not to invalidate 
   our choice of keeping fixed $\delta_\mathrm{LD}$ 
   and $\delta_\mathrm{HD}$. 
   
   We also verified 
   that the galaxy 
   density field does not evolve 
   significantly from $z=0.9$ down to 0.5 
   (i.e., we can safely compare 
   results obtained at different redshifts).  
   In fact, 
   the values of the density thresholds 
   at the 25th and 75th 
   percentile do not show a  
   dependence on $z$. 
   Moreover, by means of cosmological simulations 
   based on the Millennium Simulation
  \citep[the same used in][]{DiPorto2014} we check  
  that the PDF of the underlying matter density field
  is almost constant between $z=1$ and 0.5.
   These tests confirm that we can safely classify 
   galaxies by using the same thresholds 
   ($\delta_\mathrm{LD}$ 
   and $\delta_\mathrm{HD}$) in different $z$-bins.

  Besides that, we can estimate purity and completeness 
  of the LD and HD samples 
  by means of the HOD simulation already 
  used to test cosmic variance effects. 
  We parametrise galaxy environments as done with data,
  in both the VIPERS-like and the reference mocks,
  and classify the LD and HD environments.
  The comparison indicates that our method is not harmed 
  by the effects of the VIPERS design: in each VIPERS-like mock
  the classification is in good agreement with the one obtained
  in the reference (i.e.~working without the limitations 
  of the observational strategy).
  About 70\%   
  of galaxies for which $\delta$ is
  below the 25th (above the 75th) percentile in the reference mocks,
  remain  in the LD (HD) environment also in the 
  VIPERS-like ones. 
  For the purity, we considered the interlopers that should have 
  been associated to LD or HD (according to the 
  reference estimate) but erroneously
  fall in  the opposite environments. We find that    
  less than 8\% of low-density galaxies in the reference
  are misclassified as high-density in the VIPERS-like mocks, 
  and a similar percentage 
  of HD galaxies become LD interlopers.

\section{Volumes occupied by HD and LD galaxies}
\label{Voronoi}

In this Appendix we describe the technique 
to evaluate the comoving volumes where we recover   
 the low- and high-density regions. 
Also in this case we rely on 
the  volume-limited sample introduced 
in Sect.~\ref{Density contrast}, i.e.~those objects    
with $M_B\leqslant -20.4 - z$ 
that have been used 
to estimate the galaxy density contrast $\delta$. 
Such a sample, 
contrary to a flux-limited one,  has   
uniform characteristics from $z=0.5$ to 0.9 and  
should not introduce any redshift-dependent bias 
\citep{Cucciati2014a}.   
We already know the local  density 
contrast of these bright galaxies (Sect.~\ref{Density contrast}), 
so we can identify 
the ones that belong to LD or HD 
environments (Sect.~\ref{Environment definition}).

We fill the whole VIPERS volume 
with random particles homogeneously distributed with 
a comoving density equal to $2\,h_{70}^3\,\mathrm{Mpc}^{-3}$.  
We associate each random particle to the nearest galaxy among 
 the volume-limited sample.  Particles linked 
to LD (HD) galaxies are taken into account to 
estimate the volume occupied by the LD (or HD) regions, 
which is the fraction of particles in the specific 
environment, multiplied by the total VIPERS volume.  
Namely, this is a Monte Carlo integration  in 
comoving coordinates \citep[see e.g.][]{Weinzierl2000}. 

 We compare this estimate to an alternative technique, 
 based on the Voronoi decomposition \citep[e.g.][and references therein]{Marinoni2002}.  
 Around a chosen galaxy (belonging to the volume-limited sample), 
 a Voronoi polyhedron 
 is unambiguously defined as 
 the set of points closer to that object than to any other.   
 Once realised such a partition of the VIPERS space, we add together 
 the polyhedra of LD/HD galaxies 
 to estimate the volume of the two environments.
 This sum  overestimates the previous result 
 by $\sim20\%$, 
 because a few Voronoi 
 polyhedra exceed the effective volume of VIPERS, i.e.~they expands 
 in the VIMOS gaps. 
 On the other hand,  in the Monte Carlo integration we do not 
 deal with such a problem because 
 we can easily remove 
 random particles that fall out from the 
 spectroscopic area. We verified that, for galaxies 
 far from the survey gaps, the two techniques 
 are in excellent agreement.

\begin{figure}
    \includegraphics[width=0.99\columnwidth]{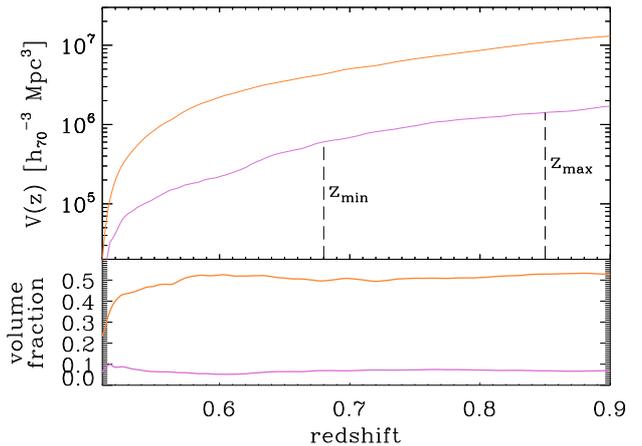}
    \caption{\textit{Upper panel:} function of the comoving volume, between redshift 0.51 and $z$, 
    filled by either HD  and LD  regions (violet and orange lines). The function is evaluated by means of 
    a Monte Carlo integration as described in the text. To find the $V_\mathrm{max}$ of a galaxy, 
    one has to consider the volume between its minimum and maximum allowed redshift 
    ($z_\mathrm{min}$ and  $z_\mathrm{min}$, see the vertical dashed lines as an example). 
    \textit{Lower panel:} the fraction of the total volume (between $z=0.5$ and the given redshift) 
    occupied by HD and LD regions (violet and orange lines). }
    \label{vmax_z}
    \end{figure}

  Once delimited low and high densities 
  in the 3-D space, we plot the LD/HD 
  volumes ($V_\mathrm{LD}$ and $V_\mathrm{HD}$)\footnote{In the following we will refer 
  to these volumes also with the general term  $V_\mathrm{env}$.}  
  enclosed between $z=0.5$  
  and a certain $z_\mathrm{up}$.
  This upper boundary   
  runs from 0.5 to 0.9 with steps of $\Delta z=0.002$.  
  That is, 
  \begin{equation}
  V_\mathrm{env}(z_\mathrm{up})=  \frac{N_\mathrm{env}(0.5,z_\mathrm{up})}
  {N(0.5,z_\mathrm{up})} V(0.5,z_\mathrm{up}) \;, 
  \end{equation}
   where $N_\mathrm{env}/N$ is the fraction of random particles -- in the 
   range $[0.5,z_\mathrm{up}]$ -- associated to the given environment,  
   while $V$ is the comoving volume of the whole survey in the same redshift slice 
   (see Fig.~\ref{vmax_z}). As said before, $V$ is computed 
   considering only the effective (i.e., spectroscopically observed) area of VIPERS and 
   the random particle outside of that are not considered.     
    We linearly 
    interpolate $V_\mathrm{env}(z_\mathrm{up})$ 
    between consecutive values of $z_\mathrm{up}$ 
    to get  a continuous function $V_\mathrm{env}(z)$, shown in the upper panel of Fig.~\ref{vmax_z}. 
    
    When computing the GSMF (Sect.~\ref{Gsmf estimate}) 
    we use $V_\mathrm{env}(z)$ to determine the 
    $V_\mathrm{max}$ volume. Each VIPERS galaxy is detectable 
    between redshift $z_\mathrm{min}$ and $z_\mathrm{max}$, 
    i.e.~the distances at which the object becomes respectively 
    brighter/fainter than the flux range of the survey. 
    In some cases $z_\mathrm{min}$ and/or $z_\mathrm{max}$  
    fall outside the  $z$-bin in which the GSMF 
    is measured. If so, we replace $z_\mathrm{min}$ ($z_\mathrm{max}$) 
    with the lower (upper) limit of the bin. 
     Once established its redshift interval of observability, 
     the $V_\mathrm{max}$  of a given galaxy is equal to    
    $V_\mathrm{env}(z_\mathrm{max}) - 
    V_\mathrm{env}(z_\mathrm{min})$, as illustrated in 
     Fig.~\ref{vmax_z}.  
     This approach is a variation of the method of \citet{Schmidt1968}, 
     accounting for the spatial segregation of 
     the sample. Indeed, the ``classical'' computation of 
     $1/V_\mathrm{max}$ is based on 
     the area of the whole survey, while here 
     we assume that galaxies contributing to the 
     LD/HD stellar mass function  cannot be observed  
     outside their environment. 
     With the exception of the first $\sim130\,h_{70}^{-1}\,\mathrm{Mpc}$ 
     along the line of sight (between $z=0.51$ and 0.55) the fraction of the total volume 
     occupied by the HD and LD structures is nearly constant, i.e.~about 7 and 
     50\% respectively (Fig.~\ref{vmax_z}, lower panel).

     This technique could be also applied to the semi-analytic mock samples 
     (Sect.~\ref{Millennium}), but in the present work we do not use it because 
     of a few systematics that make the comparison to real data more difficult. 
     One reason is that  the cosmological parameters of the Millennium simulation  
     \citep[based on WMAP1,][]{Spergel2003}  
     could be different from the ones of the observed universe. 
     In particular, the amplitude of matter fluctuations on 
    $8\,h^{-1}\,\mathrm{Mpc}$ scale should be overestimated in the simulation 
    (where $\sigma_8=0.9$) compared to more recent measurements  ($\sigma_8\simeq0.8$).  
    Also $\Omega_\Lambda$, $\Omega_\mathrm{m}$, and 
    the spectral index of the primordial perturbation field are slightly different in WMAP1 from  
    what found by WMAP9  \citep[][]{Hinshaw2013} and \textit{Planck} 
    \citep{Planck2015-13}. 
    In view of these facts,  the HD/LD thresholds in the model 
    may not agree with data. 
    Compared to VIPERS, low-density regions should be more extended, while the overdenisties 
    should be concentrated in a smaller volume, as expected in a more clustered universe. 
    These differences shall be verified in future work.

     \citet{Wang2008} investigated some consequences of varying cosmological parameters in a simulation. 
     They ran the same SAM \citep{DeLucia&Blaizot2007} several times, 
      but changing cosmology from WMAP1 to WMAP3  \citep{Spergel2007}.   
      The variations due to the new parameters mostly cancel out at 
      $z\sim0$,  while they are significant at $z\gtrsim1$.  This is especially evident by looking at the  GSMF 
      \citep[][Fig.~14]{Wang2008}, which starts to over-predict the observations already at $z=0.5$  when  
      WMAP1 parameters are assumed.  
       The luminosity function  is less affected by these systematics \citep[][Fig.~13]{Wang2008}. 
       We also notice that modifications of the galaxy formation model should 
       have a  smaller impact than cosmology on the GSMF. 

\begin{figure}
   \includegraphics[width=0.99\columnwidth]{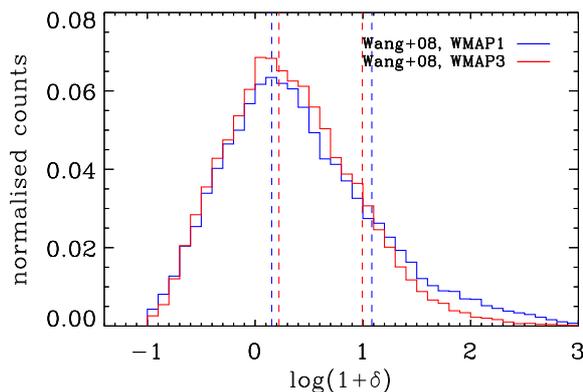}
   \caption{Distribution of $1+\delta$ in two cosmological boxes at $z=0.75$. In both simulations, galaxies 
   evolve according to \citet{DeLucia&Blaizot2007} prescriptions. The cosmological parameters used as input 
   are not the same, being taken from WMAP1 (red histogram) or WMAP3 (blue histogram). In this case, since 
   we are not restricted to projected coordinates, we evaluated the density contrast using the 5NN in the 3-d space. }
   \label{wang_hist}
   \end{figure}

       We identify low- and high-density galaxies in the   \citeauthor{Wang2008} boxes 
       ($125\,h^{-1}\,\mathrm{Mpc}$ comoving size), those based on WMAP1 
       as well as the boxes with WMAP3 cosmology. 
       We observe that  the 
       distribution of the density contrast has a higher tail at large values of $\delta$ when 
       WMAP1 is the reference. Thus, the two thresholds to divide HD and LD regions are more 
       ``extreme'' (Fig.~\ref{wang_hist}), mainly because structures form earlier  
       in the WMAP1 case.\footnote{Similar results are found by \citet{Guo2013} 
       comparing WMAP1 and WMAP7 parameters. 
       For example, at a fixed cosmic time 
       massive haloes ($>10^{12.5}\,\mathcal{M}_\odot$) 
       are more abundant with a WMAP1 cosmology \citep[][Fig.~1]{Guo2013}.}
       
       The systematic effects are even more severe when comparing our mock samples (which are based on WMAP1) to 
       observations: they are due not only to cosmology (especially $\sigma_8$) but also to differences  
       between modelled galaxies and real ones, because of both theoretical and observational limitations. 
       For example, the luminosity function predicted by \citet{DeLucia&Blaizot2007} at $z\sim0.7$ 
       has a characteristic magnitude 
       ( $M_\mathrm{B}^\star\simeq-20.5$) about 0.2\,dex brighter than the one measured in VIPERS \citep{Fritz2014}.  
       It means that  galaxies with $M_B<20.4-z$, used to to define the 5NN,  has a higher number density and 
       should trace the environment on slightly smaller scales.
       As an aside, we note that these outcomes suggest another possible use of our dataset:   
       the reconstructed volumes from observations, since they are sensitive to cosmological parameters, 
       can be used to devise a new kind of cosmological test.

\end{appendix}

\bibliography{citati-MFenv}

\end{document}